\documentclass{jfm}

\usepackage{graphicx}
\usepackage{newtxtext}
\usepackage{newtxmath}
\usepackage{natbib}
\usepackage{subfigure}
\usepackage{comment}
\usepackage{hyperref}
\hypersetup{
    colorlinks = true,
    urlcolor   = blue,
    citecolor  = black,
}

\newcommand{\RomanNumeralCaps}[1]

\title{Aerodynamic Forces on a Wing Surfing in a Two-dimensional Vortex Wake}

\author{Siyang Hao
  \corresp{\email{siyang\_hao@brown.edu}},
 Kenneth Breuer\aff{1}}

\affiliation{Center for Fluid Mechanics, School of Engineering, Brown University, Providence RI 02912 USA}

\begin{document}
\maketitle

\begin{abstract}

Inspired by the wake-surfing nature of animals, this study aims to understand the aerodynamic force variation on a wing surfing in an unsteady 2-D wake. Wind tunnel experiments were conducted using Particle Image Velocimetry (PIV) and force measurements with a fixed wing immersed in the wake of a pitching airfoil. The comparison between force and PIV measurements shows that the lift response of the surfing wing is aligned with the impingement of flow structures, and that the dependence of the cycle-averaged lift fluctuations on the upstream flapping kinematics can be scaled as a function of the reduced amplitude and reduced frequency of the flapping motion. Good collapse of the data is found, and deviations from scaling are explained in terms of the wake characteristics.  The phase-resolved lift fluctuations on the vortical wake encountering can be effectively predicted using classic unsteady aerodynamics based on measured unsteady local flow conditions (instantaneous angle of attack and speed). The theoretical predictions compare well with direct force sensor measurements.
\end{abstract}


\section{Introductions}

Wake surfing, observed in natural phenomena such as birds flying in flocks  \citep{portugal_upwash_2014, ballerini_empirical_2008, lissaman_formation_1970} or in atmospheric turbulence  \citep{laurent_turbulence_2021}, highlights the role of wake dynamics in coordinated movement, although various motivations drive this behavior.


These wakes often originate from vortex shedding from a wing undergoing unsteady motion such as pitching or plunging movements \citep{onoue_vortex_2016}, bluff body effects \citep{zhang_aerodynamics_2022}, or transverse gusts \citep{perrotta_unsteady_2017} — which commonly occur in formation flight/swimming, urban micro-unmanned aerial vehicle (MUAV) operations, and turbine arrays in wind and tidal energy contexts. As structured wakes advect downstream, they interact with objects along their path, inducing fluctuating forces and momentum. These interactions are crucial for understanding animal flight dynamics, enhancing MUAV performance, and advancing energy harvesting technologies.

\subsection{Structured wake propagation and encountering}
Structured wake propagation from pitching or plunging motions often displays coherent vortex structures in near-field advection. The nondimensional kinematic parameters — primarily Strouhal number ($St$), reduced frequency ($f^*$), and flapping amplitude($A$) — play a critical role in shaping wake patterns \citep{godoy-diana_transitions_2008,schnipper_vortex_2009}. 
Within this parameter space that is composed of these kinematic parameters, the present study focuses on cases with Strouhal numbers comparable to those commonly observed in animal flight \citep{graham_k_taylor_flying_2003}.
Studies employing similar kinematic parameters, such as, \citet{onoue_vortex_2016}, characterized vortex strength using shear layer speed from formation to growth, demonstrating the dependence of vortex strength on the kinematics of flapping motion. 
Meanwhile, \citet{turhan_coherence_2022} showed a highly coherent wake shed from a plunging airfoil across a wide range of $St$. Proper Orthogonal Decomposition (POD) analysis of the PIV results revealed that most of the energy is preserved in the first two modes, which can be further scaled as a function of reduced frequency and Strouhal number. While these studies have demonstrated that vortex wake dynamics are strongly influenced by the generative kinematics, the interactions between the wake and a downstream object remain an open question for further study.

As structured wakes advect downstream and impinge on wings in a free stream, canonical vortex–wing interaction patterns emerge, particularly when the vortex diameter is comparable to the wing chord \citep{jones_gust_2020}. A number of recent studies have examined such scenarios using different wake-generating mechanisms and wing configurations. 
\citet{qian_interaction_2022} investigated two-dimensional vortex wakes generated by a plunging foil and their interaction with swept and unswept wings, both loaded and unloaded. They provided detailed observations of the lift response as vortices approached, impinged upon, and passed the wing, and further showed—using a lumped vortex model—that the maximum lift is proportional to the circulation of the incident vortex at impingement. However, the relationship between the lift response of the tandem wing and the wake-generating kinematics was not established.
\citet{turhan_interaction_2022} partially addressed this gap by examining a wing in the wake of a plunging airfoil, showing that for a wing at zero angle of attack, the peak lift coefficient increased with the amplitude and frequency of the upstream plunging motion. This result can be extended to a dependence on the Strouhal number ($St$), since the Strouhal number—defined as the ratio of plunging velocity to freestream velocity—can itself be interpreted as a nondimensional measure of wake amplitude. However, the peak lift may not be the most appropriate metric for quantifying the intensity of lift oscillations on the surfing wing; a clearer scaling collapse between wake generation and wake interaction could likely be achieved using alternative measures.

Collectively, these studies indicate a strong correlation between the aerodynamic fluctuations experienced by downstream bodies and both the characteristics of the upstream wake structures and the kinematics of their generation.
Building on this foundation, the present work examines how generating kinematics, impinging vortex structures, and the resulting lift fluctuation intensity influence a downstream wing under wake-surfing conditions.

\subsection{Predictive models}

To model and predict the forces experienced during wake interactions, quasi-steady assumptions are often employed for their balance between simplicity and effectiveness. For example, \citet{sane_aerodynamic_2002} used a quasi-steady blade element approach to successfully capture the aerodynamics of wing rotation in insect flapping. However, as \citet{perrotta_unsteady_2017} demonstrated, when moving to a higher Reynolds number, quasi-steady models tend to significantly overpredict aerodynamic forces under such conditions. This is because quasi-steady models inherently approximate only the instantaneous aerodynamic state, neglecting the time history or memory effects that are fundamental to high Reynolds number unsteady flows. In addition, the inviscid assumption in thin airfoil theory fails to capture viscous phenomena such as flow separation, which become increasingly important at higher Reynolds numbers ($Re$). The vortex-impulse-based method \citep{saffman_vortex_1993}, like the lumped vortex model used by \citet{qian_interaction_2022} to predict lift responding on the encountering of vortex wake, demonstrated the potential for time-resolved predictions. Although effective, its application typically requires global flow-field information, such as vortex location and strength, which limits its suitability for practical prediction; thus, it will not be much discussed in the present work.

As an alternative, to incorporate unsteady aerodynamic effects, analytical models have been developed based on local instantaneous flow conditions with varying emphasis on the treatment of fluctuations.
Theodorsen's function  \citep{bisplinghoff_aeroelasticity_1996} provides a foundational model for predicting the lift of an airfoil in simple harmonic motion. Extending from the harmonic motion of the airfoil to periodic transverse gust profiles,  \citet{von_karman_airfoil_1938} proposed a broken-line analogy which led to Sear's function. 

Further extensions of these theories include the indicial lift response, such as Wagner's function for step-pitching motions and K\"ussner's function for sharp-edge gusts. These functions can be
derived from Theodorsen's function through Fourier synthesis with different initial conditions  \citep{dowell_modern_2022}. (However, this is not the only approach, nor does it reflect the historical development of these theories.) 

Based on the measured unsteady local flow conditions, such as instantaneous angle of attack and speed, these classic unsteady aerodynamic theories can be used to predict the lift forces' response to the wake encountering, and interpret experimental observations. 
However, these theories have limitations arising from their underlying assumptions of linearity, thin airfoil theory, inviscid flow, and the wake's evolution being unaffected by the presence of the airfoil  \citep{bisplinghoff_aeroelasticity_1996}.

In recent work, \citet{turhan_interaction_2022} demonstrated a pathway for quantifying the amplitude of lift fluctuations on a wing positioned in the wake of a plunging airfoil using Sears’ analytical model. However, their study did not provide time- or phase-resolved predictive capability, leaving the dynamic prediction of the unsteady lift response unresolved.
\citet{perrotta_unsteady_2017} modeled transverse gust encounters for a flat-plate wing using a convolution-based K\"ussner function, which outperformed quasi-steady models. Notably, the use of the indicial response method enabled time-resolved predictions of gust encounters. However, the authors questioned the applicability of the indicial approach for accurately predicting vortex-induced force oscillations. Developing a model that can incorporate local vorticity features while leveraging unsteady aerodynamic theory through the indicial method remains a significant challenge.

Although these studies offer only partial descriptions of the force oscillation prediction during wake encountering, they provide valuable insight and form the foundation of the present study, which extends the comparison to time-resolved experimental data across multiple classical unsteady aerodynamic models.
\subsection{The structure of this work}
In this work, wind tunnel experiments were conducted using Particle Image Velocimetry (PIV) and force measurements with a fixed wing immersed in the wake of a pitching airfoil.
We demonstrate (i) the dependence between wake generation, advection, and interaction with the surfing wing, and (ii) how the theory can be adapted to account for coherent vortical wake structures—extending the classic linear framework to quasi-linear and viscous flow conditions—supported by our experimental data.

In the following section, we introduce our experimental setup and the time-resolved lift prediction method (Section~\ref{method}).  In Section~\ref{force-flow}, we examine the relationship between lift and flow patterns by comparing the phase-averaged vorticity field with synchronized force measurements. Section~\ref{cycle-lift} analyzes cycle-level fluctuations in the lift coefficient using spectral narrow-band integration. In Section~\ref{prediction}, we develop a predictive method based on unsteady aerodynamics and local flow information (instantaneous angle of attack and flow speed) to capture time-resolved lift variations. 
Finally, we summarize the key findings in Section~\ref{summary}.

\section{Methods}
\label{method}
\subsection{Experimental setup}

\begin{figure*}
\centering
\includegraphics[width=4in]{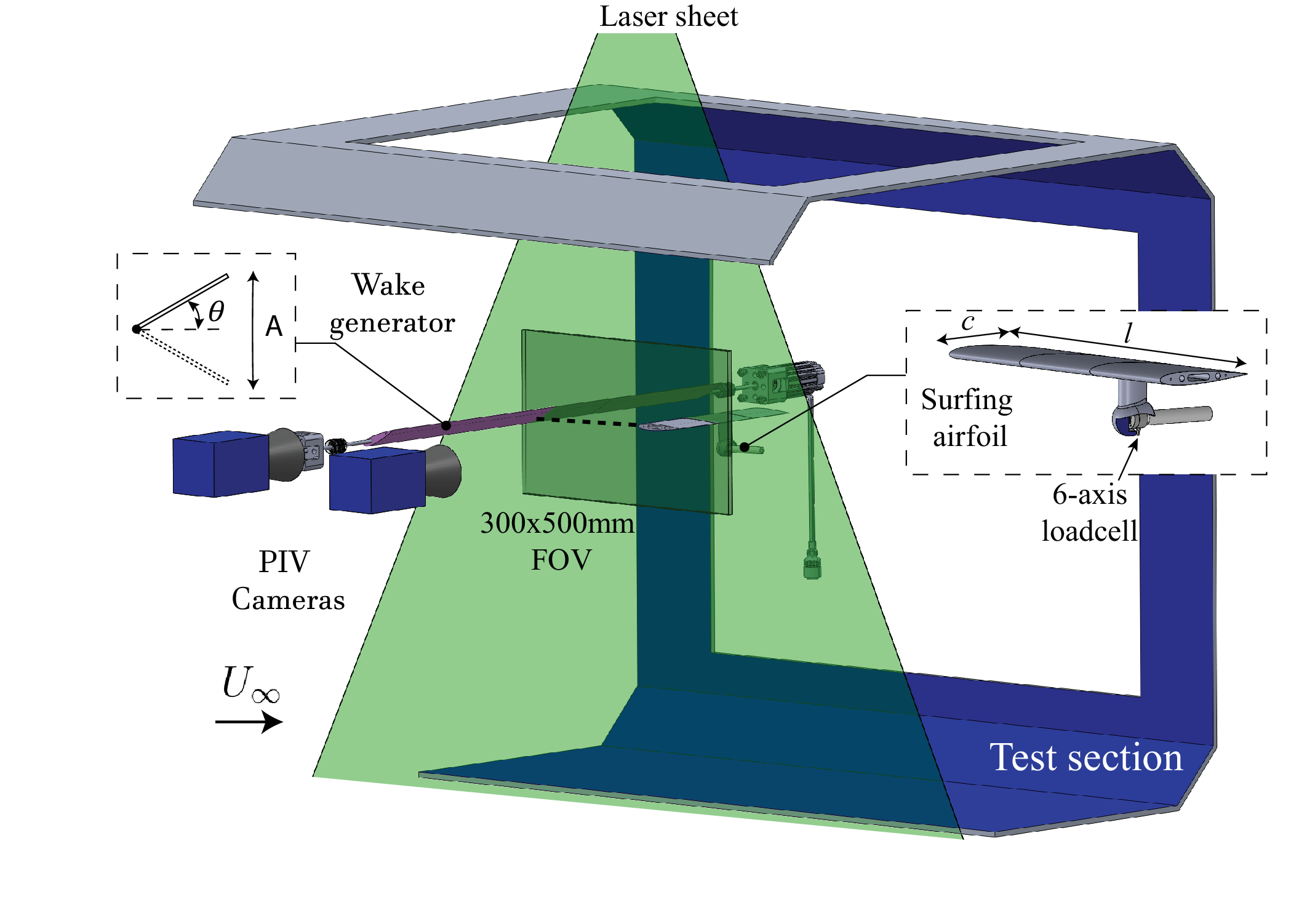}
\caption{Experimental setup in wind tunnel. The motorized wake generator is installed upstream, with a flat plate spanning the test section pitching sinusoidally. An NACA0012 airfoil is aligned with the center of the wake and mounted to loadcell through a streamlined support. Both the surfing wing and the wake generator are included in a 300x500 streamwise field of view (FOV) of the PIV, taken by the two side-by-side PIV cameras.
}
\label{fig:2DEngExp}
\end{figure*}

As shown in figure~\ref{fig:2DEngExp}, experiments were conducted in the Animal Flight and Aero-Mechanics Wind Tunnel at Brown University  \citep{breuer_design_2022}, with the test section of 1.2m by 1.2m. A servo-motor (Hudson, Teknic Inc., NY) driven  wake generator consisting of a flat plate with a chord length of $c = 100 mm$, spanning the entire width of the test section was installed in the upstream portion of the test section.  The plate executed sinusoidal pitching about its leading edge, with prescribed frequencies ranging from $2$ to $8\,\text{Hz}$ and nondimensional amplitudes ($A/c$) from $0.2$ to $0.7$, following the motion  
\[
\theta(t) = \frac{A}{2} \sin(2\pi f t).
\]
Initially, the pitching profile is created in Matlab (Mathworks, MA), generating 1024 path points per cycle. These points are then buffered to the motion controller (DMC-4060, Galil, CA), which converts them into servo motion using polynomial smoothing. The position error is monitored by the encoder of the servo motor and is maintained at less than 1\% of the maximum amplitude.  

A NACA 0012 wing (chord $c = 100\text{mm}$, span $l = 300 \text{mm}$) was positioned at the same height as the wake generator, where the largest amplitude of lift fluctuations occurs according to  \citet{turhan_coherence_2022}, 350 mm downstream from the flapping plate. The wing was mounted on a 6 DoF force/torque load cell (Nano-17, ATI, NC) via a streamlined support, which reduces interference with the load cell.
The force responses were measured at 5000 Hz over 100 flapping cycles and then phase-averaged. 

To study the vortex dynamics, Particle Image Velocimetry (PIV) was used to measure the flow field around the wake generator and the wing. The flow was seeded with $1 \mu m$ diameter DEHS aerosol generated using a Laskin nozzle. The mid-span plane  of the flow was illuminated using a double-pulse Nd:YLF laser (527 nm, 40 mJ/pulse DM40, Photonics, NY) with in-house collimating optics and LaVision sheet optics.
Two high-speed cameras (Nova R2, Photron Inc, CA) were positioned side-by-side outside the test section and used to acquire images at $100$ times the flapping frequency with a $500mm \times 300mm$ field of view covering both the wake-generator and the surfing airfoil.
The recorded PIV images were processed using the DaVis software (v10, LaVision, two passes at 32 × 32 pixels, two passes at 24 × 24 pixels, both with 75\% overlap) to calculate the velocity vectors. The DaVis software estimates the uncertainty in velocity vectors to be below 3\% of the maximum velocity. Finally, the velocity fields obtained from the two cameras were stitched together to form a $3c \times 5c$ field of view.

\begin{figure}
\centering
    \includegraphics[width= 0.5\textwidth]{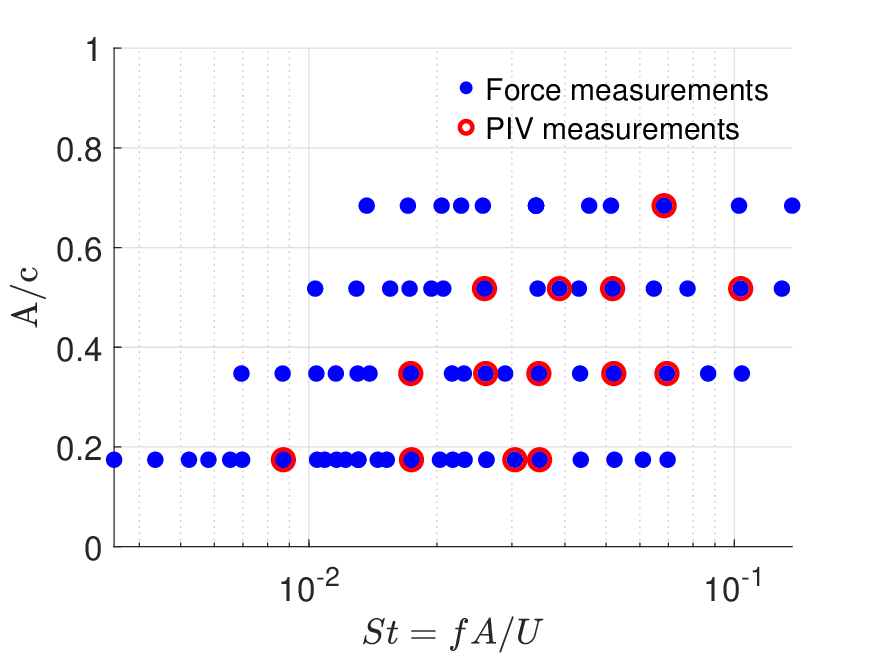}
\caption{ The parameter sets of tested cases
}
\label{fig:dataset}
\end{figure}

In these experiments, the free-stream velocity ranged from $U = 2$ m/s to 10 m/s, while the flapping frequency varied between $f = 2$ Hz and 8 Hz and the pitching amplitude varied between $A/c=0.17$ to $0.67$. This resulted in non-dimensional Strouhal numbers, \(St = fA/U\), ranging from 0.004 to 0.14, and reduced pitching amplitudes, \(A/c\), ranging from 0.17 to 0.68. As shown in figure~\ref{fig:dataset}, a total of 90 force measurement cases were conducted, each covering 100 flapping cycles, along with 14 synchronized PIV measurement cases highlighted by red markers.

The entire experiment is controlled by a Matlab script. A custom-built control box  \citep{siyang_hao_triggerbox_2021} was implemented to receive and output TTL signals, which synchronize the PIV, motion control, wind tunnel, and data acquisition with microsecond precision.

\subsection{Time-resolved lift prediction based on local flow field}

Here, we review the key theoretical methods to predict the aerodynamic forces on the downstream wing.  Starting with the simplest theory - steady, thin airfoil theory - the lift coefficient of a two-dimensional wing can be expressed as \citep{anderson_fundamentals_2017}:
\begin{equation} C_{L, \text{QS}, 2D} = 2\pi \alpha_{\text{ins}} , \end{equation}
where $\alpha_{\text{ins}}$ is the instantaneous angle of attack.   A simple enhancement of this, which accounts for  horizontal and vertical mean flow fluctuations, $u$ and $v$ respectively, as well as time varying angle of attack result in:

\begin{equation} 
\label{QuasiCL} 
C_{L, \text{QS},2D}(t) = 2\pi \alpha_{\text{ins}}(t) \sqrt{\frac{u^2(t) + v^2(t)}{U_{\infty}^2 }}  
\end{equation}

\subsubsection{Wagner and K\"ussner function with Duhamel Integral}

The Wagner function \citep{dowell_modern_2022} can be used to model the unsteady aerodynamic forces resulting from instantaneous changes in the pitch angle of an airfoil. Originally, Wagner's solution accounts for an indicial pitching motion of the airfoil. In our case, however, the wing is stationary while encountering the gust; the effect of the vortical gust encounter is considered as an indicial change in angle of attack without wing motion. Therefore, the added mass term in the original formulation is not included in our discussion. In this instance, the unsteady lift coefficient (excluding the added mass terms) derived from the Wagner function can be expressed as the product of the quasi-steady lift coefficient, considering the local speed, and the ``Wagner'' function:
\begin{equation}
    C_{L,W,2D} = C_{L,QS,2D}(t) \Phi(\tau)
\end{equation}
%
%
where:
\begin{align*}
    \tau &= \frac{2tU_{\infty}}{c}  .
\end{align*}
The Wagner function, $\Phi(\tau)$, converges at 1 with an infinitely large $\tau$ as the unsteady effect takes time to develop into quasi-steady.  An approximation of $\Phi(\tau)$ is given by  \citep{fung_introduction_2008}:
\begin{equation}
    \Phi(\tau)  \approx \frac{2+\tau}{4+\tau} . 
\end{equation}
%
%
%
The Duhamel integral (see the Appendix) can be introduced to superpose the continuous time-varying $\alpha_{ins}$ by integrating a series of small step responses to take the time history into account (as discussed by  \citet{karman_mathematical_1940}). Thus, the time-varying lift coefficient can be predicted using the Wagner function as:
\begin{equation}
    C_{L,W,2D}(t) = C_{L,QS}(0)\Phi(\tau) + \int _{0}^{\tau}\Phi(\tau-\tau_0)\frac{dC_{L,QS}}{d\tau_0}d\tau_0 \ .
    \label{ConvWag}
\end{equation}

Lastly, we can account for finite aspect ratio wings \citep{perrotta_unsteady_2017}, resulting in the unsteady lift coefficient expressed as 
\begin{equation}
 C_{L,W}(t) = \left[C_{L,QS,2D}(0)\Phi(\tau) + \int _{0}^{\tau}\Phi(\tau-\tau_0)\frac{dC_{L,QS,2D}}{d\tau_0}d\tau_0\right] \frac{AR}{AR+2} \ .
\end{equation}

The K\"ussner function models the lift response of a wing encountering a sharp-edged transverse gust, and following the same procedure outlined above, it can be applied as a modification of the quasi-steady lift coefficient: 
\begin{equation}
    C_{L,K,2D}  = C_{L,QS,2D}(t) \Psi(\tau) \ .
\end{equation}
An approximation of the K\"ussner function is given by  \citep{bisplinghoff_aeroelasticity_1996}:
\begin{equation}
    \Psi(\tau) \approx \frac{\tau^2+\tau}{\tau^2 + 2.82\tau + 0.8}
\end{equation}
and, as with the Wagner function, can be applied to a time-varying gust profile using the Duhamel integral:
\begin{equation}
    C_{L,K}(t) = \left[C_{L,QS,2D}(0)\Psi(\tau) + \int _{0}^{\tau}\Psi(\tau-\tau_0)\frac{dC_{L,QS,2D}}{d\tau_0}d\tau_0\right] \frac{AR}{AR+2} \ .
    \label{ConvKuss}
\end{equation}

\subsubsection{Sears Function with Fourier Integral}
Lastly, for an oscillatory flow with a given reduced frequency \( k = \pi c f/U_{\infty}\), the responding unsteady aerodynamic forces in the frequency domain can be modeled using the Sears function \citep{dowell_modern_2022}: 
\begin{equation}
    S(k) = \left[J_0(k) - iJ_1(k)\right]C(k) + iJ_1(k)
\end{equation}
where \( C(k) \) is Theodorsen’s function, and \( J_j (k) \) (for \( j = 0, 1, \dots \)) are Bessel functions of the first kind. With the modification by \( S(k) \), the unsteady lift coefficient is given by:

\begin{equation}
    C_{L,S,2D}(k) = C_{L,QS,2D}(k) \cdot S(k)
\end{equation}
Where the $C_{L,QS,2D}(k)$ is the Fourier transformation of the time-varying lift coefficient, $C_{L,QS,2D}(t)$: 
\begin{equation}
    C_{L,QS,2D}(k) =\mathcal{F}\{C_{L,QS,2D}(t)\} .
\end{equation}  
Combining these, with the finite aspect ratio correction results in:
\begin{equation}
    C_{L,S}(t) = \mathcal{F}^{-1}\{\mathcal{F}\{C_{L,QS,2D}(t)\} \cdot S(k)\}\frac{AR}{AR+2} . 
\end{equation}

\section{Results and Discussion}

\subsection{Lift force fluctuations and wake structures}
\label{force-flow}
\begin{figure*}
\centering \includegraphics[width= \textwidth]{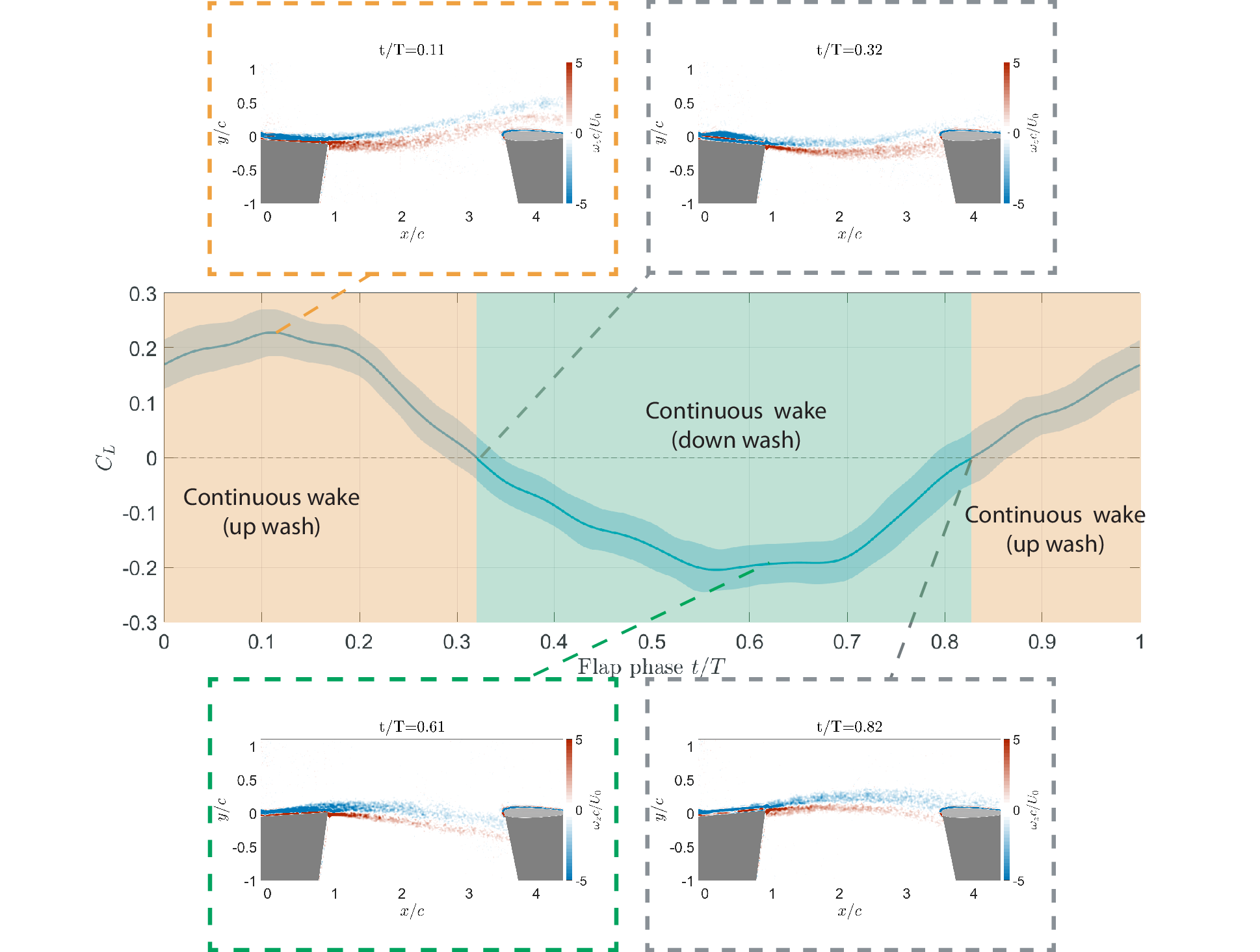}
\caption{Unsteady coefficient of lift (center) and four instantaneous vorticity distributions at $t/T = 0.11, 0.32, 0.61$ and 0.82. $\text{Re} = 42\text{k}$, $\text{St} = 0.017$, $A/c = 0.2$.  At these conditions we observe a continuous vortex sheet shed from wake generator and impinging on the surfing wing.}
\label{fig:surferPIVLow}
\end{figure*}

\begin{figure*}
\centering
\includegraphics[width=\textwidth]{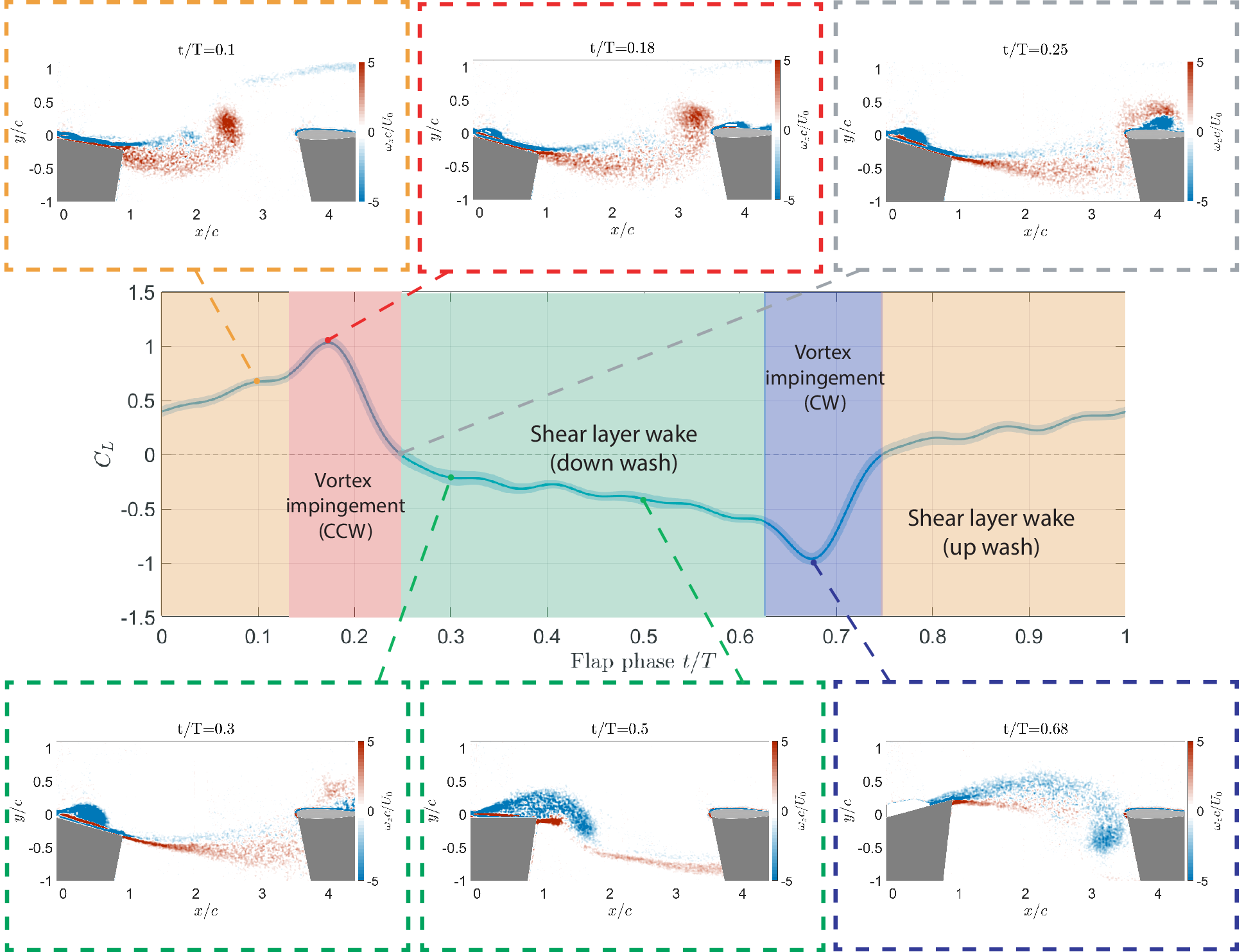}
\caption{Unsteady coefficient of lift (center) and six instantaneous vorticity distributions at $t/T = 0.10, 0.18, 0.25, 0.30, 0.50$ and 0.68.
At $\text{Re} = 42\text{k}$, $\text{St} = 0.052$, $A/c = 0.5$.  At these conditions we observe discrete vortex shedding from the flapping wing, and impingement on the surfing wing.}
\label{fig:surferPIVHigh}
\end{figure*}

The relation between the lift force experienced by the surfing wing and the flow pattern can be revealed by comparing the synchronized and phase-averaged force measurements and the PIV. The direct force measurement is presented as the lift coefficient: 
\begin{equation}
  C_L = \frac{L}{1/2 \rho U_{\infty}^2 c l}  , 
\end{equation}
which is shown for two Strouhal numbers in figure~\ref{fig:surferPIVLow} and figure~\ref{fig:surferPIVHigh}.
$t/T = 0$ corresponds to the beginning of the sinusoidal motion of the wake generator when the plate passes through the neutral position in a downward motion. 
The $z$-vorticity field (perpendicular to the measurement plane) is non-dimensionalized as $\omega_z c / U_{\infty}$ and shown at different times during the stroke cycle.

The flow field and forces resulting from a low St number case at $Re=42k$, $St = 0.017$, $A / c = 0.2$ are shown in figure~\ref{fig:surferPIVLow}. We see a continuous wake shed from the pitching plate, resulting in a smooth, gradually changing lift coefficient experienced by the surfing wing. Due to the motion of the pitching plate, the wake shifts up and down.  As it passes over the top surface of the surfing wing, it induces a positive lift coefficient (shaded in an orange color). The lift coefficient shifts to negative during the half-stroke when the wake moves to the lower surface of the surfer (shaded in green). 

As the Strouhal number increases, discrete vortices are shed from the pitching plate, and influence the force response on the surfing airfoil. The flow field and forces resulting from flapping at a higher Strouhal number at $Re=42k$ -  $ St = 0.052$, $A / c = 0.5$  - are shown in figure~\ref{fig:surferPIVHigh}. A pair of leading edge vortices are shed during each stroke of the pitching motion: a counter-clockwise (CCW) vortex (shown in red in figure~\ref{fig:surferPIVHigh}) from the downstroke and a clockwise (CW) vortex (blue) from the upstroke.  For example, a CW leading-edge vortex forms and grows on the upper surface of the wake generator ($t/T = 0.1-0.3$) and detaches at $t/T = 0.5$. The vortex advects downstream ($t/T >= 0.5$). A mirrored formation process of CCW vortex can be anticipated half-cycle away, but the PIV image is partially blocked by the shadow cast by the wing. 

As the CCW vortex advects downstream, the surfing wing begins to experience an increase in the lift coefficient (as the period highlighted in orange in figure~\ref{fig:surferPIVHigh}). The lift coefficient reaches a positive peak at the moment of impingement ($t/T = 0.18$), when a separation bubble starts to form on the upper surface. As the separation bubble moves downstream with the CCW vortex and detaches ($t/T \approx 0.25$), the lift coefficient rapidly drops back to a neutral state. When the separation moves away and the shear layer wake shifts to the lower side of the surfing foil, the lift becomes negative. This entire advection process is also mirrored during the second half of the cycle: A negative lift peak occurs one half-cycle away as the CW vortex is encountered ($t/T = 0.68$).

These results are consistent with the findings of  \citet{qian_interaction_2022}, who report on a wing in the wake of a plunging airfoil, and find that the aerodynamic forces induced by a single discrete vortex show a gradual increase in the lift coefficient as the vortex approaches, a rapid increase upon impingement, and a sharp drop with the initiation of the separation bubble.\citet{turhan_interaction_2022} observed a slightly different lift profile when a wing surfs in the vortex street of a plunging airfoil. Their findings indicated a more rapid transition between positive and negative lift, accompanied by a dip around the time of impingement. 
The reduced frequency ($f^* = \pi c f/U_{\infty}$) in our experiments falls between those two studies. We observed counterclockwise (CCW) and clockwise (CW) vortex impingement occurring in a sparser sequence, resulting in small interactions between them. Each impingement is more comparable to that of the single vortex considered by \citet{qian_interaction_2022}.

\begin{figure*}
  \subfigure[$C_L$ deviate with flapping amplitude]{
    \includegraphics[width= 0.49\textwidth]{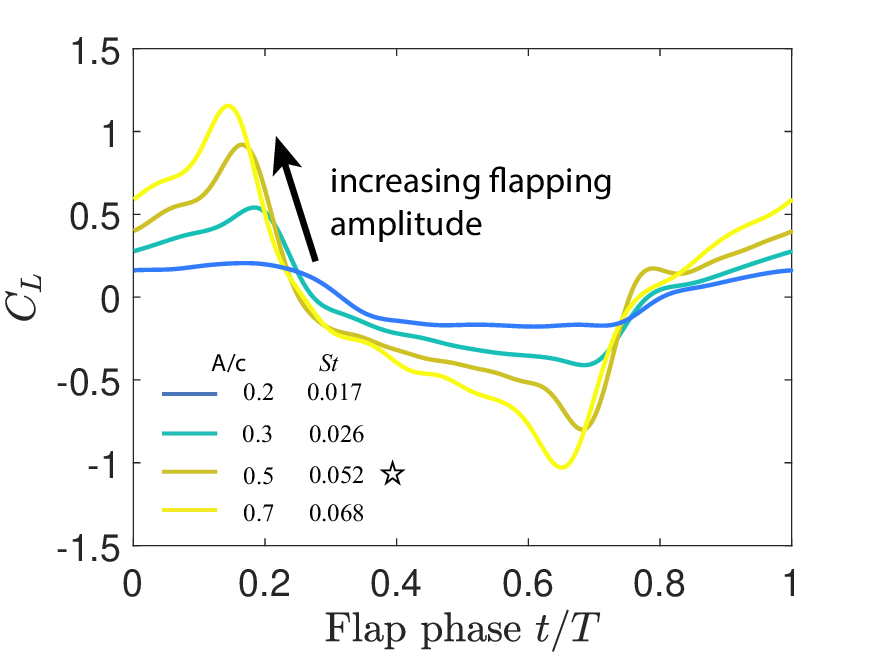}
    }
    \subfigure[$C_L$ deviate with flapping frequency]{
    \includegraphics[width= 0.49\textwidth]{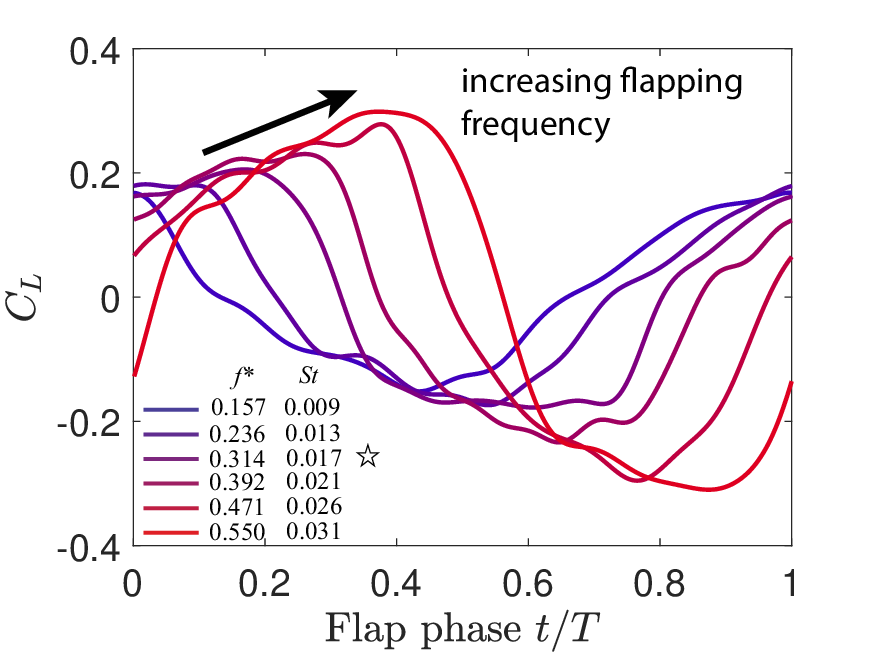}
    }
\caption{(a) As the flapping frequency of the wake generator increases (with a constant free-stream velocity $U_{\infty} = 4m/s$ and flapping amplitude $f_0 = 4Hz$), the lift coefficient fluctuations increase with a slight phase shift forward.The case shown in figure \ref{fig:surferPIVHigh} is marked by star symbol.
(b) As the flapping frequency of the wake generator increases (with a fixed free-stream velocity $U_{\infty} = 4m/s$ and flapping amplitude $A/c = 0.2$), the lift fluctuations also increase with a phase shift backward.The case shown in figure \ref{fig:surferPIVLow} is marked by star symbol. 
}
\label{fig:CLdev}
\end{figure*}

The synchronized PIV and force measurements show a good correlation between the lift coefficient fluctuations and the wake structures. Hence, the force deviation may reflect subtle variations in wake patterns induced by changes in wing kinematics, such as Strouhal number and reduced frequency. For a fixed free-stream velocity of $U_{\infty} = 4m/s$ and fixed flapping frequency of $f_0 = 4Hz$ ($f^* = 0.31$), increases in flapping amplitude result in a corresponding increase in measured lift, and the peak due to the discrete vortex impingement becomes more distinctive (Figure~\ref{fig:CLdev}a). The $C_L$ profile also exhibits a slight shift towards an earlier phase. Similarly, with an increase in the flapping frequency of the wake generator, but keeping the free-stream velocity fixed at $U_{\infty} = 4m/s$ and the flapping amplitude fixed at $A/c = 0.2$, the lift fluctuations increase, but this time we see a phase delay (figure~\ref{fig:CLdev}b). 

For the deviation of the peak lift coefficient, \citet{qian_interaction_2022} showed a proportional relationship between the vortex circulation ($\Gamma/U_{\infty}c$) and the peak lift fluctuations ($\Delta C_{L,\max}$). It is therefore not surprising that, in our case, $\Delta C_{L,\max}$ increases with the reduced amplitude and frequency of the generating motion, as both contribute to enhancing the wake circulation.  As for the phase lag, this can be attributed to the combined time required for the vortex to form and detach from the pitching plate ($t_{\text{sep}}$), and for it to advect downstream to the surfing wing ($t_{\text{adv}}$):
\[
t_{imp} = t_{sep} + t_{adv} .
\]

\begin{figure*}
  \subfigure[Scaled CCW vortex impingement time deviate with reduced flapping frequency]{
    \includegraphics[width= 0.5\textwidth]{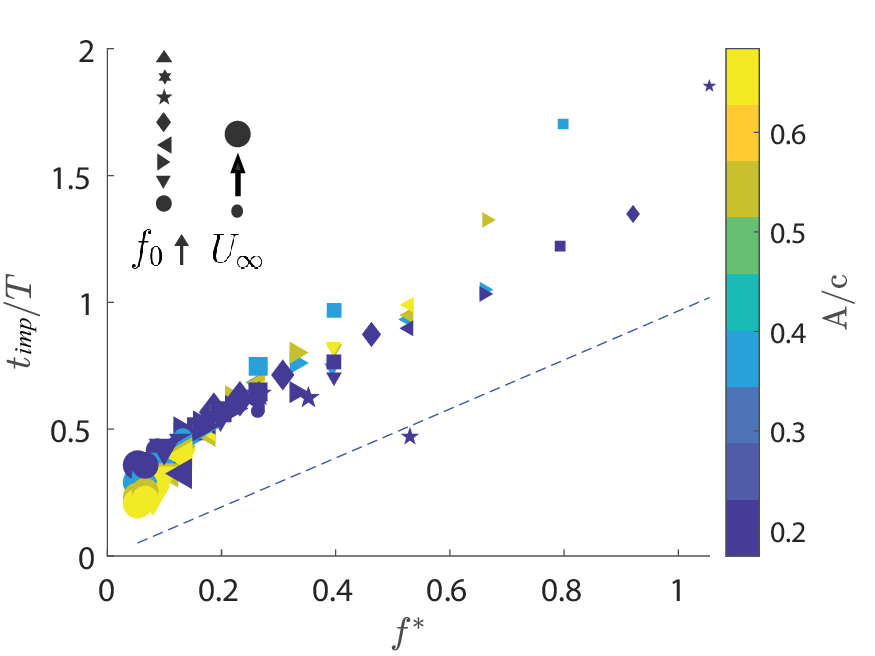}
    }
    \subfigure[Scaled CCW vortex detaching time deviate with flapping $St$ number]{
    \includegraphics[width= 0.5\textwidth]{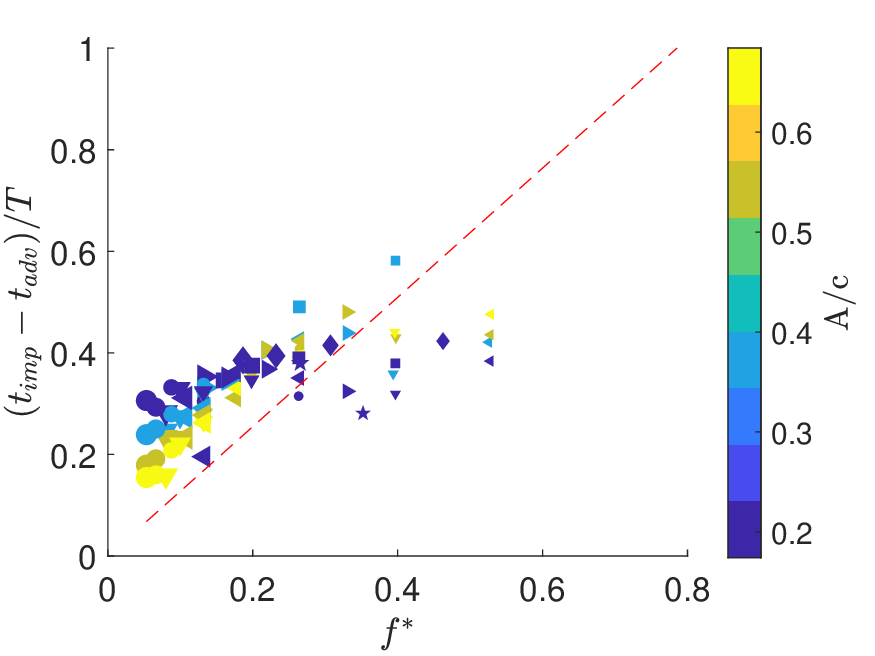}
    }
\caption{The scaled impingement time, $t_{imp}$, of CCW vortices, tracked by the positive peak of force measurement, is shown in (a). The blue dashed line represents the advective time based on free-stream velocity and the distance from the wake generator to the surfing airfoil. In (b), when the calculated advective time is subtracted from the total impingement time, the delay from the beginning of the flapping motion to the moment the vortex leaves the trailing edge of the wake generator can be shown as a function of the reduced frequency $f^*$ for each flapping amplitude.
}
\label{fig:phase_delay}
\end{figure*}

The moment of vortex impingement, $t_{imp}$ can be identified from the peak in the $C_L$ profile (\ref{fig:CLdev}).  This is shown, for the counterclockwise (CCW) vortex in Figure~\ref{fig:phase_delay}a, against the reduced frequency for a range of freestream velocities and flapping frequencies.

%
%
The advection distance, $d$ - the distance between the wake generator and the surfing wing - is fixed, and if we assume that the vortex moves at the freestream velocity, we can write the scaled advection time, $t_{adv}/T$ as
\[ 
t_{adv}/T = f d/U_{\infty} = f^*\frac{d}{c \pi} . 
\] 
This is shown by the blue dashed line in figure~\ref{fig:phase_delay}a.  Subtracting the advection time, $t_{adv}$, from the total impingement time, $t_{imp}$, we can isolate the vortex formation and detachment time, $t_{sep}$, shown as a function of the reduced frequency $f^*$, in figure~\ref{fig:phase_delay}b. 
In general, the vortex formation and detachment tends to rise with $f^*$. At low values of $f^*$ the collapse of the data is pretty good, but does have a weak dependence on amplitude (color of the symbols). The scaling breaks down for higher values of $f^*$,  although the data here is more sparse and this range of $f^*$ is associated with low values of $U_{\infty}$. 

As discussed by several previous studies \citep[e.g.][]{dabiri_optimal_2009,onoue_vortex_2016,gehrke_phenomenology_2021}, the formation and shedding process is strongly governed by the characteristic shear-layer ``feeding'' velocity, $U_{SL}$ - the characteristic velocity of the flow that feeds the leading edge vortex, and a limiting ``formation time" \citep{dabiri_optimal_2009}:
\(
t^* = t U_{SL} / D,
\) 
where $D$ is a characteristic length scale. To adapt this scaling to our configuration, the estimated dimensionless formation time takes the form  
\[
t^* = \frac{t_{\text{imp}} - t_{\text{adv}}}{T} \propto \frac{t^* U_{\infty}}{\pi U_{SL}} f^* , 
\]  
which is shown as the red dashed line in figure~\ref{fig:phase_delay}b, using $D = c, U_{\infty}/U_{SL} = 1$, and $t^* = 4$ \citep{dabiri_optimal_2009}. This simple scaling argument matches the amplitude and trend of the measurements very well. 
The noted dependence on amplitude can be accounted for the fact that as $A/c$ increases, one would expect the shear layer velocity, $U_{SL}$, to increase as the speed around the leading edge increases, thus suppressing $t_{sep}$.


\subsection{Spectral analysis of the lift fluctuation}
\label{cycle-lift}
\begin{figure*}
\centering
    \includegraphics[width= .6\textwidth]{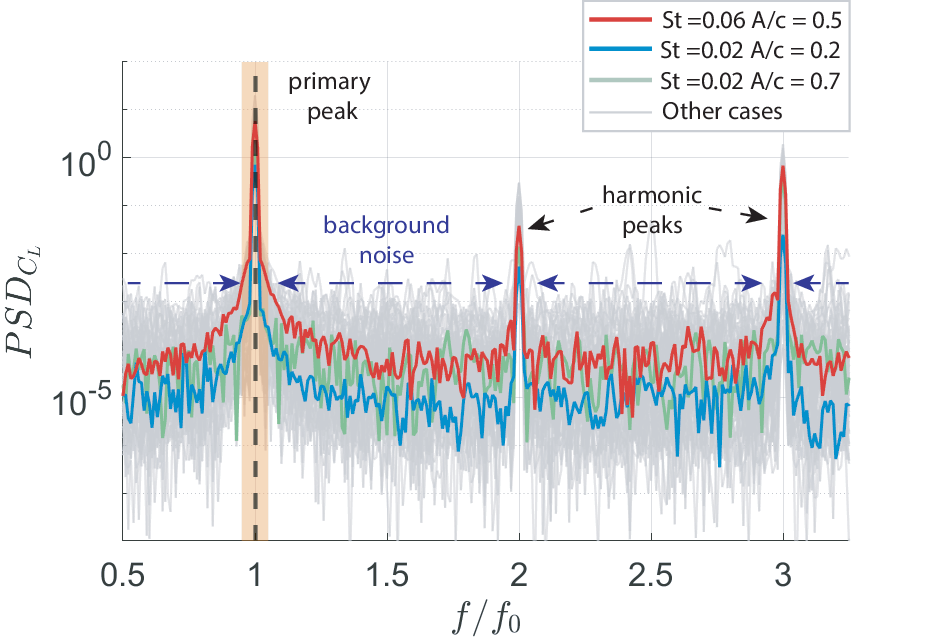}
\caption{Power spectral density of the lift coefficient for all measurements, with the case of $Re = 42k, St = 0.06$, and $A/c = 0.5$ (also shown in figure~\ref{fig:surferPIVHigh};
$Re = 42k, St = 0.02$, and $A/c = 0.2$ (also shown in figure~\ref{fig:surferPIVLow}); $Re = 105k, St = 0.02$, and $A/c = 0.7$ highlighted. 
}
\label{fig:surferPSD}
\end{figure*}
The response of the surfing airfoil can be further analyzed in the frequency domain. figure~\ref{fig:surferPSD} shows the Power Spectral Density (PSD) of the measured lift coefficients from 90 trials, representing a wide range of frequencies, amplitudes, and wind speeds. A few cases are highlighted as examples.  The $x$-axis depicts the response frequency, scaled by the flapping frequency of the wake generator, $f/f_0$.

The red highlighted case of $Re = 42k, St = 0.052$, and $A/c = 0.5$ is the same as shown in figure~\ref{fig:surferPIVHigh} and represents the case of a discrete vortex wake. The blue highlighted case,  $Re = 42k, St = 0.017$, and $A/c = 0.2$,  is the same case as shown in figure~\ref{fig:surferPIVLow}, and represents the continuous wake interaction mode. The green highlighted case has the same Strouhal number, but is at a higher Reynolds number ($Re = 105k$, and $A/c = 0.7$).  

As indicated by the vertical black dashed line, the spectrum peaks at 1 and its harmonics, confirming that the fluctuation in the lift coefficient is primarily driven by the flapping frequency. Besides the primary peak, we observe that the third harmonic's peak intensity exceeds that of the second harmonic, which is consistent with  \citet{turhan_interaction_2022} and expected since the waveform of the lift coefficient is odd - with positive and negative lift peaks associated with the up and downstrokes of the pitching airfoil respectively. 

The intensity of the primary peak can be further quantified using a narrow band integration over the primary peak area (highlighted by the orange area in figure~\ref{fig:surferPSD}): 
\begin{equation}
     C_{L,P1}= \sqrt{\int_{0.95}^{1.05}PSD_{C_L}d\frac{f}{f_0}}
\end{equation}
The square root is for consistency, as the infinite integral over the frequency domain corresponds to the variance of the lift coefficient according to Parseval's theorem. 

\begin{figure*}

\subfigure[]{
    \includegraphics[width= 0.5\textwidth]{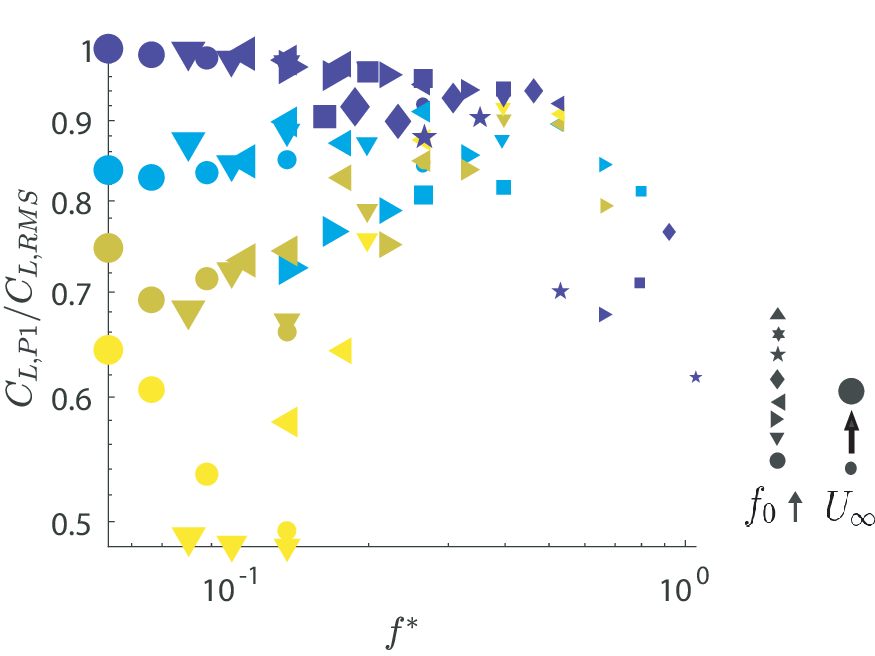}
    }
\subfigure[]{
    \includegraphics[width= 0.5\textwidth]{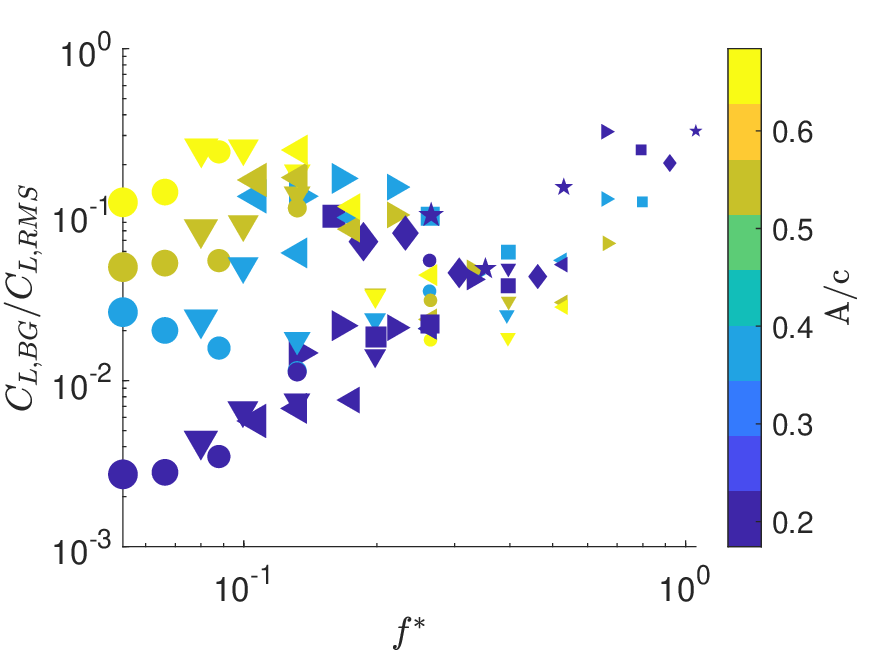}
    }
\caption{
(a) Partition of fluctuation energy around the primary peak $C_{L,\text{P1}}$ within the total fluctuation $C_{L,\text{RMS}}$.
(b) Partition of background fluctuation energy $C_{L,\text{BG}}$ within the total fluctuation $C_{L,\text{RMS}}$. The driving frequency of the wake generator, \(f_0\), is represented by different markers, the free-stream velocity, \(U_{\infty}\), is indicated by the size of the markers, and the reduced amplitude, \(A/c\), of the wake-generating pitching motion is color-coded.
}
\label{fig:P1factor}
\end{figure*}
Figure~\ref{fig:P1factor}a shows the ratio of the narrow-band integration around the primary peak to the total energy of the lift fluctuation, (the root mean square (RMS) of the lift coefficient). For all cases, most of the energy - $47\%$ to $99\%$ - is concentrated within the range $f/f_0 = [0.95, 1.05]$ (figure~\ref{fig:P1factor}a). With the continuous wake sheet generated at low flapping amplitudes, shown by the blue symbols in figure~\ref{fig:P1factor}, the lift coefficient fluctuation profile is highly sinusoidal (as seen in figure~\ref{fig:surferPIVLow}) and consequently, most of the energy is concentrated around the primary peak at the flapping frequency. As the flapping amplitude of the wake generator increases, the lift response associated with discrete vortex impingement strengthens, (figure~\ref{fig:CLdev});  the surfer's lift profile becomes less sinusoidal, leading to more energy being distributed into the harmonic peaks of the PSD (figure~\ref{fig:surferPSD}) and the portion of energy in the primary peak decreases, (figure~\ref{fig:P1factor}a).

However, as the reduced frequency, $f^*$, of flapping increases, the scaled impingement time (represented by the red and blue regions in figure~\ref{fig:surferPIVHigh}, defined as $t_{\text{imp}}/T = f c/U_{\infty} = f^*/\pi$) extends, causing the lift profile to revert to a more sinusoidal shape (as shown in figure~\ref{fig:CLdev}), which increases the fraction of energy in the primary peak. In figure~\ref{fig:P1factor}a, for each flapping amplitude except $A/c = 0.2$, the portion of narrow-band integration increases with $f^*$. The low velocity cases (high $f^*$, small symbols) tend to break these trends and are more randomly distributed.

For $A/c = 0.2$, as $f^*$ increases, there is a noticeable increase in the energy distribution of the background noise (as defined in figure~\ref{fig:surferPSD}, integral of the spectrum excluding the first 5 harmonic peaks) shown in figure~\ref{fig:P1factor}b. Consequently, unlike other amplitudes, where the harmonic peaks account for a larger portion of the total energy and the noise portion remains almost constant, the portion of energy in the primary peak decreases with $f^*$. Additionally, as shown in figure~\ref{fig:P1factor}b, as the flow experiences greater disturbances due to higher reduced frequency and increased flapping amplitude of the wake generator, the flow field becomes more turbulent. Consequently, the integration of background noise into the power spectrum, $C_{L,\text{BG}}$, shows an increased energy distribution.

\begin{figure*}
\subfigure[$C_{L,P1}$ deviates with $f^*$ and $A/c$]{
    \includegraphics[width= 0.5\textwidth]{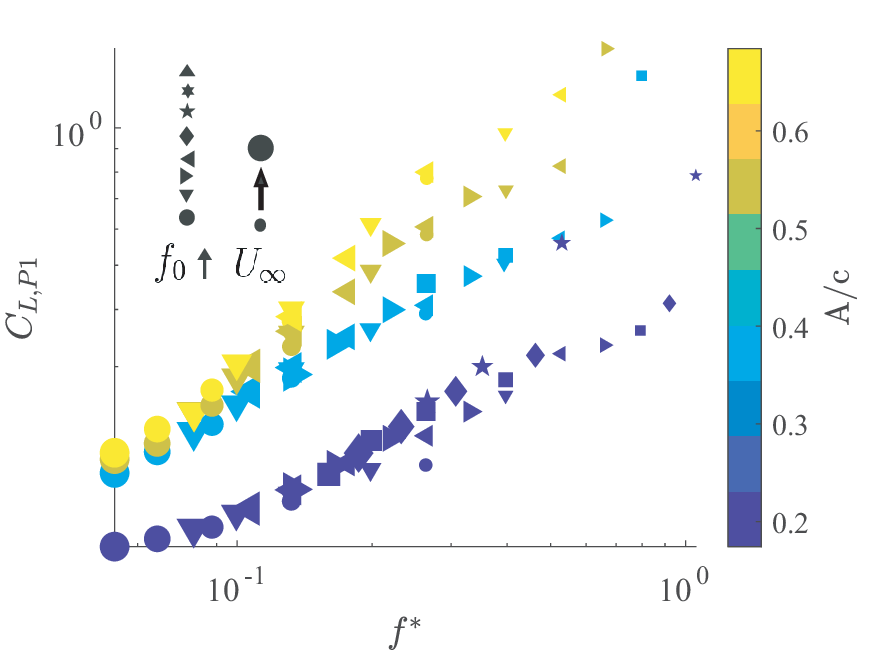}
    }
\subfigure[$C_{L,P1}$ deviates with $St$]{
    \includegraphics[width= 0.5\textwidth]{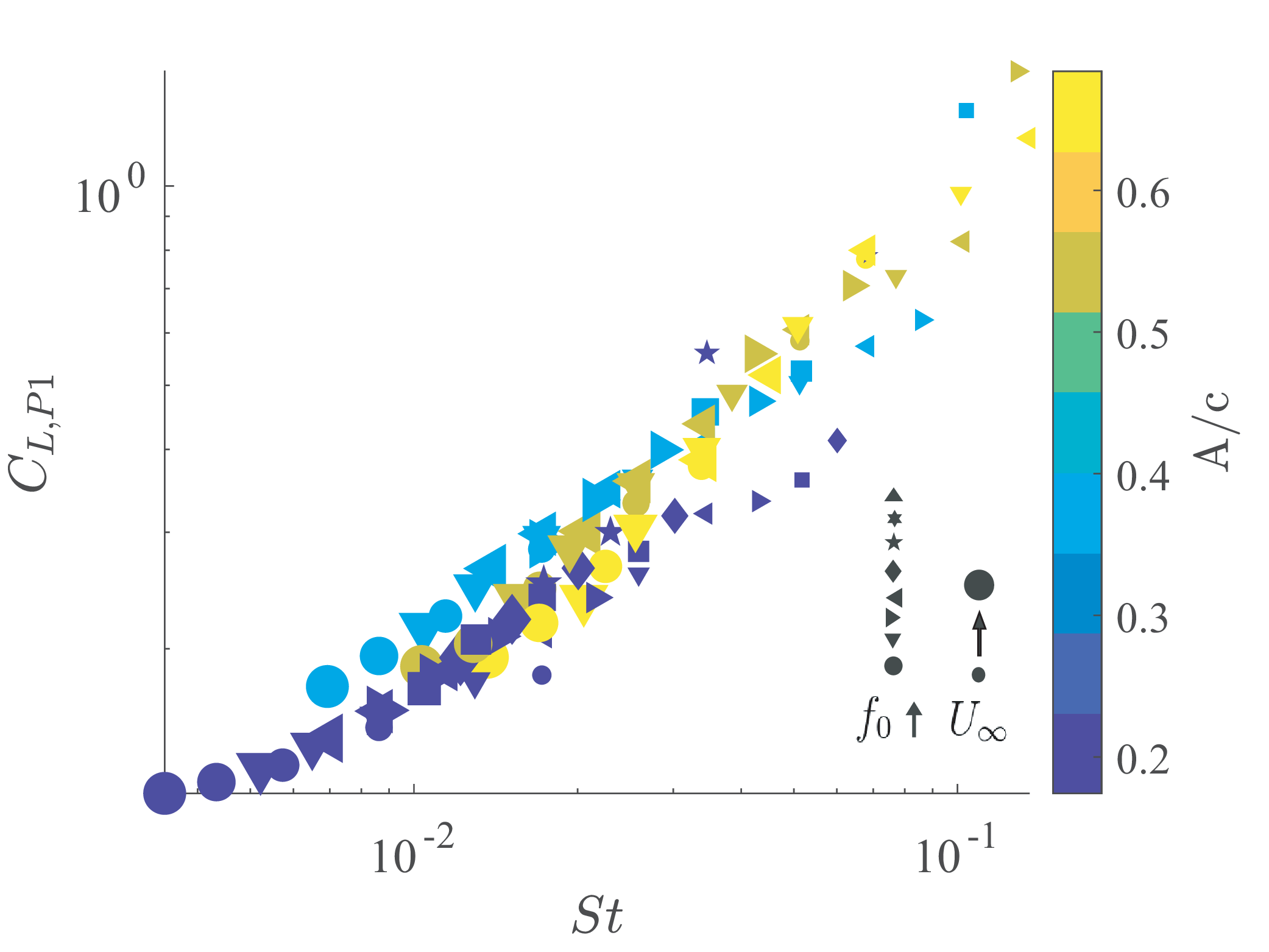}
    }
\caption{Fluctuation of lift coefficient quantified by narrow band integration is shown as a function of reduced frequency, $f^* =\pi f c/U_{\infty}$.  
}
\label{fig:P1}
\end{figure*}

The fluctuation of the lift coefficient, quantified by the narrow-band integration \(C_{L,P1}\), is shown as a function of reduced frequency \(f^* = \pi f c / U_{\infty}\) in figure~\ref{fig:P1}a. \(C_{L,P1}\) increases with \(f^*\), and the slope increases with the reduced amplitude \(A/c\) of the wake generator. As \(A/c\) increases, \(C_{L,P1}\) shifts abruptly to a higher level. The distinct gap between \(A/c < 0.2\) and \(A/c > 0.4\) arises from the wake pattern shifting from a continuous wake sheet to discrete vortices. To account for this dependence on both \(A/c\) and \(f^*\), \(C_{L,P1}\) is plotted as a function of the Strouhal number \(St = f A / U_{\infty}\) which gives a better collapse of the data (figure~\ref{fig:P1}b).
\citep{turhan_interaction_2022} also shows that the maximum value of the lift coefficient fluctuation, \(\Delta C'_{L,\text{max}}\), on the wing can also be scaled by \(St\). Qualitatively, \(\Delta C'_{L,\text{max}}\) is somewhat comparable to \(C_{L,P1}\) in our case. Quantitatively however, \(C_{L,P1}\) provides better information on the entire wake encountering process while maintaining robustness over varying waveforms across a wide test domain from $St= 0.004$ to $0.2$. As a result, we observe a clear scaling behavior with \(St\).

\subsection{Time-resolved lift predictions and comparison with theoretical models}
\label{prediction}

\begin{figure}
\subfigure[]{
    \includegraphics[width= .5\textwidth]{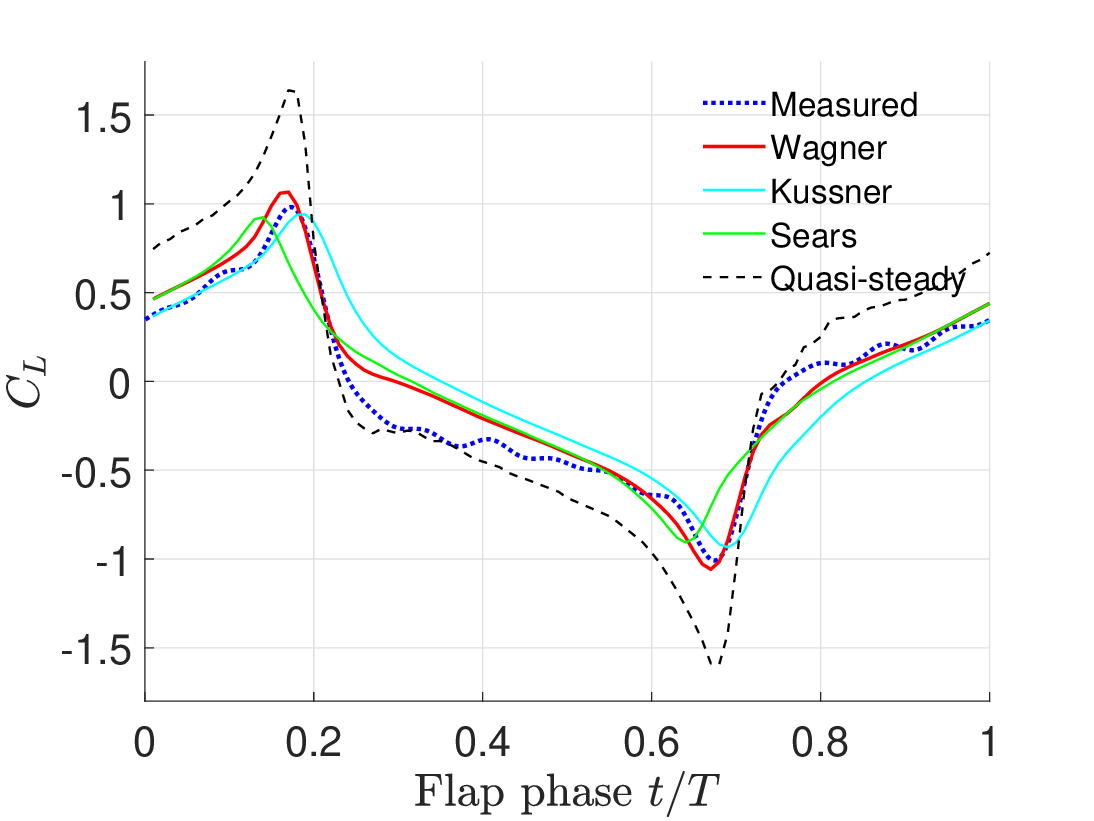}
    }
    \subfigure[]{
    \includegraphics[width= .5\textwidth]{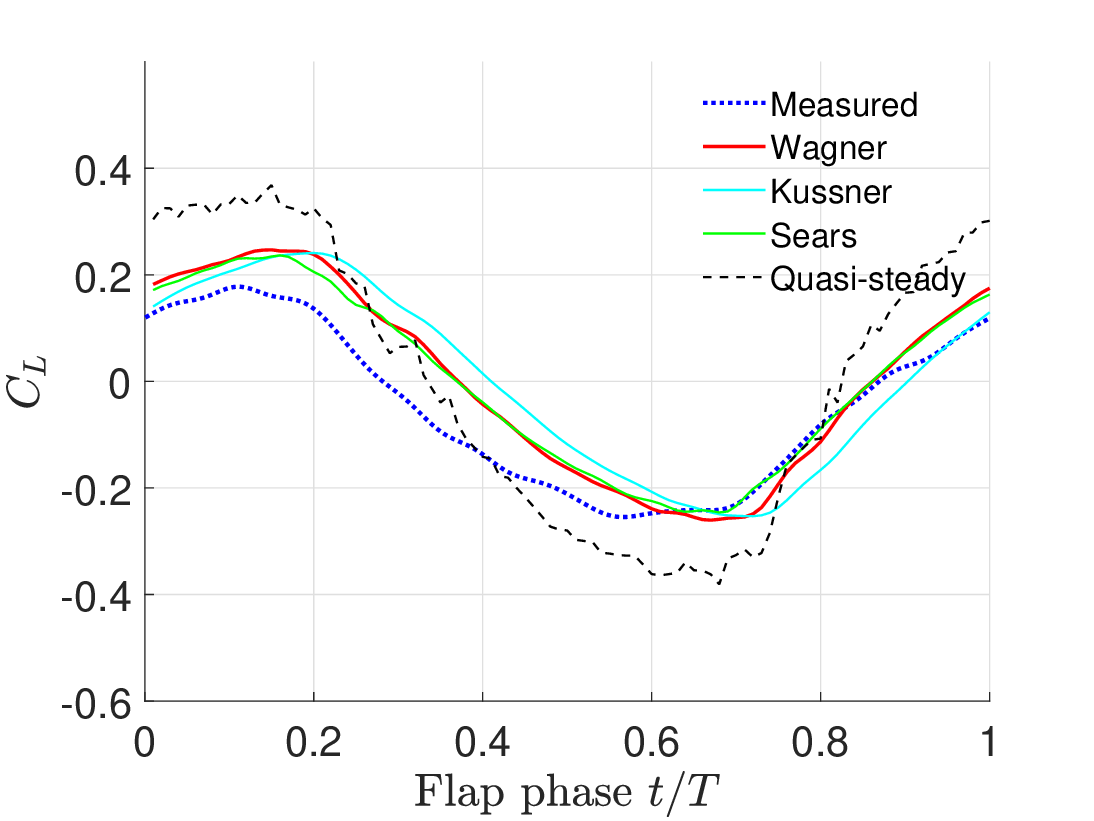}
    }
\caption{A comparison between direct force measurement, quasi-steady and unsteady approaches of a) the case of vortex impingement at Re=42k St = 0.052 A/c =0.5  and b) Re=42k, St = 0.017, A/c =0.2. 
}
\label{fig:UnsteadyResults}
\end{figure}
A comparison between direct force measurement, quasi-steady and unsteady theoretical predictions is shown in figure~\ref{fig:UnsteadyResults}(a) at $Re=42k, St = 0.052, A/c =0.5$  and (b) $Re=42k, St = 0.017, A/c =0.2$.
The phase-averaged direct force measurements from the load cell are shown in a blue dotted line. 

The quasi-steady prediction (shown as the black dashed line in figure~\ref{fig:UnsteadyResults}) is based on the experimentally determined instantaneous angle of attack, sampled at the leading edge of the surfing airfoil, used with steady thin airfoil theory with the aspect ratio correction, (Eqn~\ref{QuasiCL}). Perhaps not surprisingly, the agreement with measurement is not very good; it severely overestimates the lift.
As  \citet{perrotta_unsteady_2017} also observed when applying quasi-steady thin airfoil theory to the encounter with a transverse gust, similar overestimations occur mainly due to the breakdown of the theory under these unsteady conditions.  \citet{qian_interaction_2022} also observed a significantly smaller value when comparing the lift contribution from the encounter with a traveling vortex gust---though not time-resolved---to the classic \(2\pi\alpha\) theory.

Unlike the quasi-steady method, unsteady methods incorporate the flow history into the prediction, thus  providing a more accurate prediction. 
In the time domain, indicial functions such as the Wagner function (in red) and the K\"ussner function (in cyan) are implemented with the Duhamel integral to extend step responses into time-varying responses (Eqs.~\ref{ConvWag} and \ref{ConvKuss}). The fundamental difference between the Wagner and K\"ussner functions lies in how the wing encounters the flow structure---either all at once or progressively along the chord  \citep{jones_gust_2020}.
For purely transverse gust encounters, the convolution-based K\"ussner model can yield good agreement with aerodynamic force predictions, using an effective angle of attack across various gust profiles—particularly when the gust ratio \(GR = V/U_{\infty}\) is less than 0.5  \citep{andreu-angulo_effect_2020, biler_experimental_2019}, where the vorticity contribution remains low.
In our vortex encountering cases, since the vortex scale is comparable to the wing chord, the wing experiences a Lamb-Oseen-type velocity profile  \citep{saffman_vortex_1993}, which includes both transverse and streamwise velocity components.
The K\"ussner-function-based method exhibits a greater phase delay because it is derived for sharp-edged gusts with a rigid gust shear layer. In such scenarios, the gust encounter progresses gradually along the chord, leading to an extended rise time and slightly reduced amplitude. \citet{sedky_physics_2022} successfully employed a combined Wagner-K\"ussner model to separately account for the unsteady effects of transverse gust encounters and wing pitching. In the present case, the vortex encounter is more analogous to a sudden pitch motion (without the added mass term) because (1) the rotational velocity features are better represented, and (2) the wing encounters the wake as a whole entity.
 
In the frequency domain, the Sears function, applied with Fourier synthesis, (shown in green) has good estimation of amplitude but lacks phase accuracy in the high $St$ case in figure~\ref{fig:UnsteadyResults}a.
This may be because of the limitation of the sampling rate of our PIV dataset, which results in lower resolution in the FFT of the implementation of the Sear's function.  At the lower $St$ case, shown in figure~\ref{fig:UnsteadyResults}b, where the gust profile is more sinusoidal and the discrete vortex is not distinct, the Sear's approach on the frequency domain is comparable to the other two approaches on the time domain. All in all, in comparing the unsteady approaches, the method using the Wagner function offers the closest prediction to the directly measured data.

There is an overestimation after the vortex impingement for all of the unsteady prediction methods since 1) the separation has occurred at this moment which breaks the no-separation assumption and 2) the gust profile gets deformed by the presence of the surfing wing. 
To better account for the post-impingement separation and to address the effect of leading-edge vortex (LEV) shedding, low-order models like the Leading-Edge Suction Parameter-based Discrete Vortex Method (LDVM)  \citep{sureshbabu_theoretical_2021} could be employed in future work.

\begin{figure}

    \subfigure[Sears approach]{
    \includegraphics[width= 0.5\textwidth]{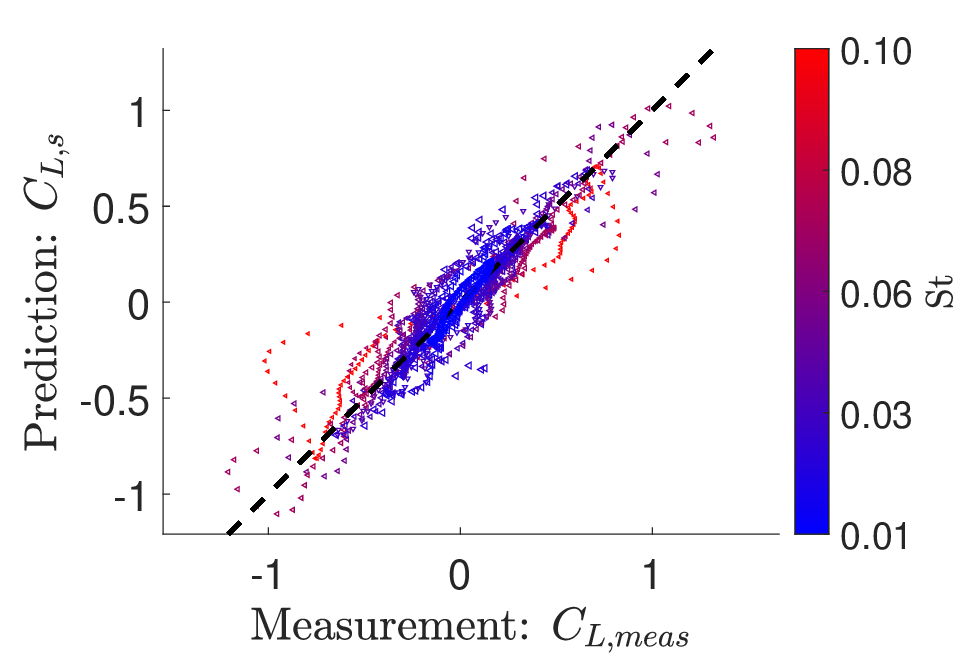}
    }
\subfigure[Wagner approach]{
    \includegraphics[width= 0.5\textwidth]{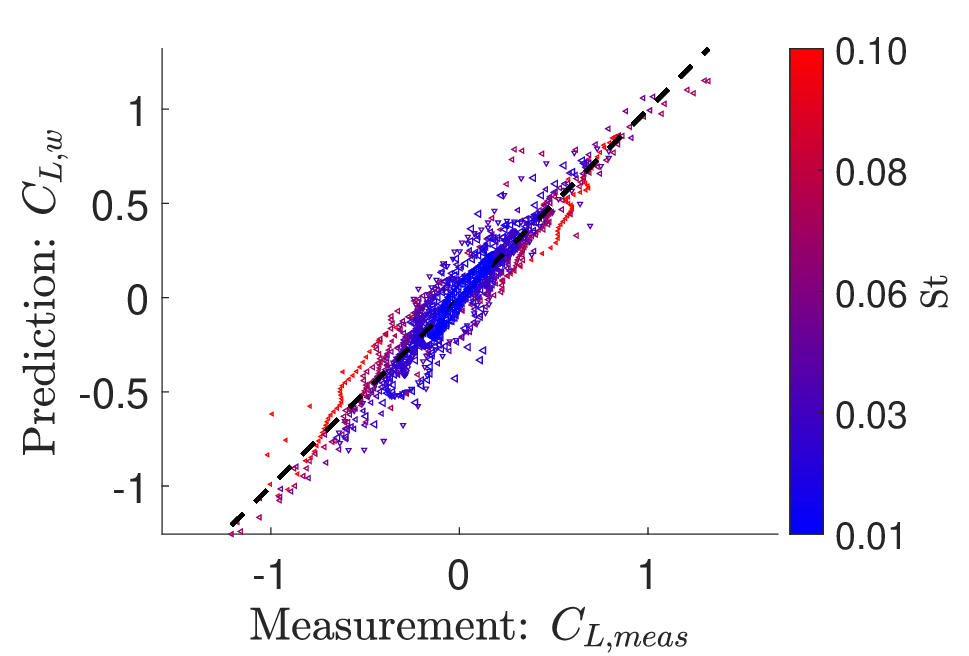}
    }

\caption{Frame-by-frame comparison of predicted vs. measured lift coefficient for various Strouhal numbers.
}
\label{fig:error_line}
\end{figure}

In figure~\ref{fig:error_line}, the predicted lift and measured lift are compared frame-by-frame across all flow conditions and for 14 phase-averaged PIV cases - a total of 1400 instants in time, including the cases shown in figure~\ref{fig:UnsteadyResults}. This comparison evaluates each phase-averaged frame for various cases with different Strouhal numbers, color-coded for distinction, showing the differences between the calculated and measured lift coefficients. The black dashed line represents a perfect match between the measured and calculated lift coefficients.

As shown in figure~\ref{fig:error_line}a, the Sears approach demonstrates higher accuracy at lower $St$ values (blue dots). However, as $St$ increases and discrete vortices are shed from the pitching wing (red dots), the calculations tend to miss the moment of vortex impingement, as shown in figure~\ref{fig:UnsteadyResults}a. In contrast, as shown in figure~\ref{fig:error_line}b, Wagner’s approach more accurately captures the effects of vortex wake impingement at higher $St$ values, and there are fewer large deviations from the 1:1 line.
Note the error shown in figure~\ref{fig:error_line} stems from both phase and amplitude; further discussion regarding these two factors can be found in the appendix.

\section{Conclusions}
\label{summary}

In this work, we have studied the lift response of a fixed wing immersed in the 2D parallel vortex wake generated by an upstream pitching plate. The force response of the surfing wing is aligned with the impingement of flow structures, and the vortex wake's generation and advection from different flapping kinematics are revealed by deviations in the lift profile. Spectrum analysis shows that the response frequency is locked to the wake-generating frequency. The narrow-band integration around the primary peak, where most of the energy is concentrated, indicates that cycle-level lift fluctuations scale with wake characteristics, particularly the Strouhal number, reduced frequency, and reduced amplitude of the wake generator's kinematics.

In the second part of this work, we applied a predictive method by extending classic linear unsteady aerodynamics to account for coherent structured wake interactions, supported by our experimental data. The theoretical predictions, based on localized flow information such as instantaneous angle of attack and speed, compare well with direct force sensor measurements. The comparisons were primarily made between the Wagner and K\"ussner approaches in the time domain and the Sears approach in the frequency domain. Due to the nature of the wake interaction in this case, the Wagner approach provides the best fit in terms of both phase and amplitude accuracy.

Even in the best case, with the Wagner approach, the predictions tend to underestimate the force during impingement and overestimate the force when separation bubbles occur on the leading edge of the wing. The primary sources of error are: 1) massive separation on and after vortex impingement and 2) deformation of the wake. These factors violate the assumptions of thin airfoil theory and non-separation on which unsteady aerodynamics are based. These errors are a concern and will be addressed in follow-up studies.

In general, this work contributes to a comprehensive understanding of wing interaction with advecting vortex gusts and bridges the upstream kinematics of the wake generator to the cycle-level fluctuations of the downstream surfing wing, as well as the flow conditions and phase-resolved lift variations.

\backsection[Acknowledgements]
{We would like to acknowledge Jared Ramirez and Raul Ayala who helped with the initial experiments during their summer research at Brown University. We appreciate the valuable discussions with Pedro Costa-Ormonde and other members of the Breuer Lab.
Many of the ideas discussed in this work were inspired by the research and insightful discussions with Ismet Gursul, whose influence continues to resonate with our approach to unsteady aerodynamics. We dedicate this work to his memory.}
\backsection[Funding]{This work was supported by NSF IOS-1930924.  }
\backsection[Declaration of interests]{The authors report no conflict of interest.}
\backsection[Author ORCIDs]{ S. Hao, http://orcid.org/0000-0001-5012-3605; K. Breuer, http://orcid.org/0000-0002-5122-2231}
\clearpage

\section*{Appendix}
\subsection*{Definition and sampling of instantaneous angle of attack}
The unsteady lift is estimated primarily from the instantaneous angle of attack. For this, we evaluate two sampling methods based on PIV data, both yielding comparable results. Additionally, examining the effect of different sampling geometries provides meaningful guidelines for the placement of sensors, like pressure or flow speed transducers, to acquire flow data using methods other than PIV.
\begin{figure*}
\includegraphics[width= \textwidth]{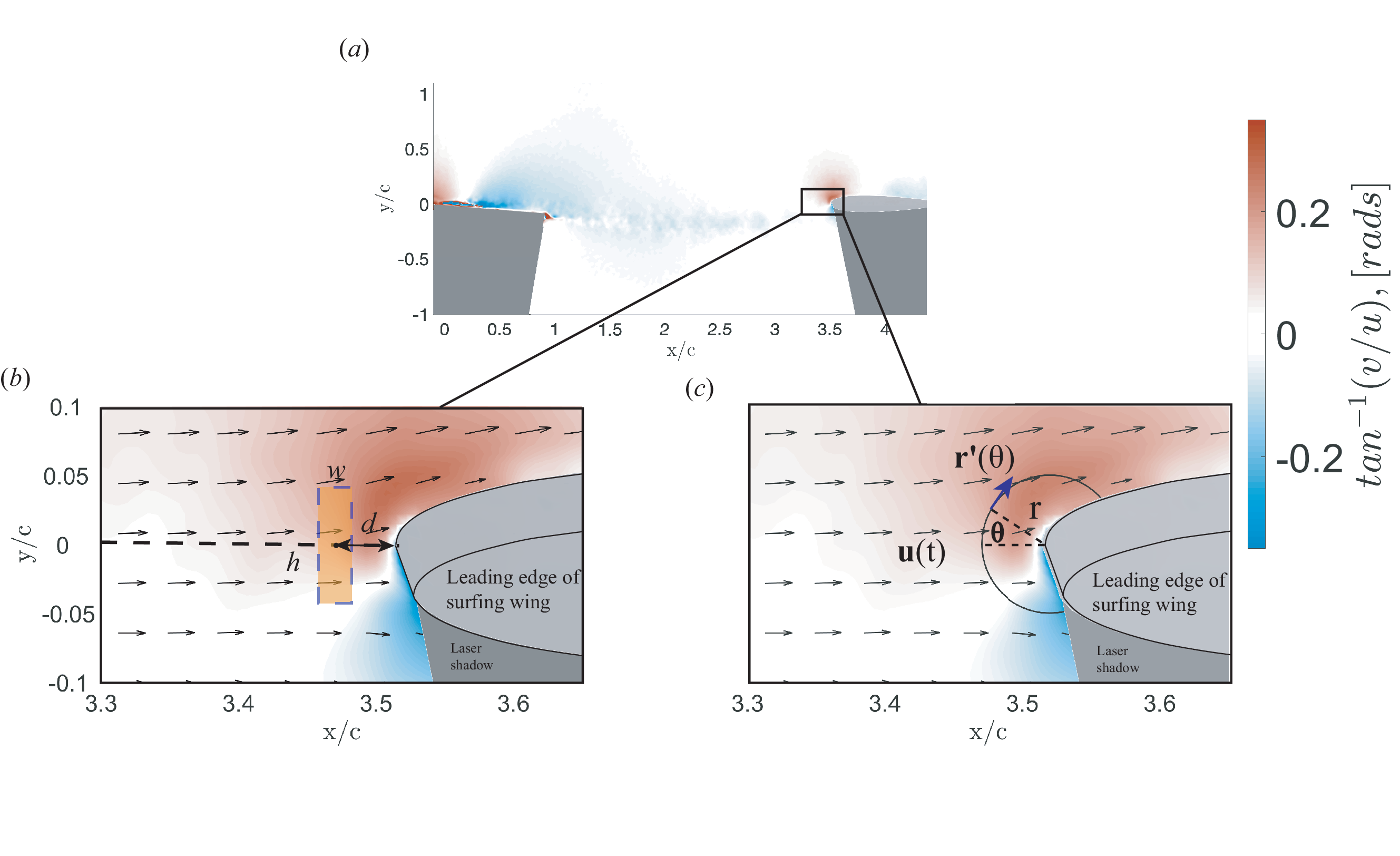}
\caption{An example of instantaneous angle of attack sampling, (a), is presented with a zoomed in figures, (b) and (c), on the leading edge of the surfing airfoil (masked by gray). The velocity field is measured using PIV, with the absolute speed $U/U_{\infty}$ displayed in color and the velocity vectors shown in a quiver plot. When using area averaging, as shown in (b), a rectangular shaped sampling area of width $w$ and height $h$ is centered $d$ from the leading edge of the surfing airfoil.
When using streamline interception, as shown in (c), a sampling circle is centered at the leading edge of the surfing airfoil, with a radius $r$. The velocity vectors $\textbf{u}(t)$ on the sampling circle are obtained by interpolating the PIV data.
}
\label{fig:LEAOA4}
\end{figure*}
\subsubsection*{Area averaging method}

In figure~\ref{fig:LEAOA4}, a schematic of the area averaging method is presented, with a zoomed-in view of the leading edge of the surfing airfoil (masked in light gray). The velocity field is measured using PIV, with the absolute speed $U/U_{\infty}$ displayed in color and the velocity vectors represented in a quiver plot. A rectangular sampling area, with width $w$ and height $h$, is centered at a distance $d$ from the leading edge of the surfing airfoil. The instantaneous angle of attack (AOA), $\alpha_{ins,avg} = \tan^{-1}(v/u)$, is calculated using the velocity data at each PIV grid point within the sampling area and the overall instantaneous AOA is then determined as the average over the sampling area.
\subsubsection*{Streamline interception method}
To better capture the geometric and flow characteristics near the leading edge, we also employed a sampling method based on local streamline orientation. As shown in figure \ref{fig:LEAOA4}c, a sampling circle is centered at the leading edge of the surfing airfoil with a radius $r$. The velocity vectors $\textbf{u}(t)$ on the sampling circle are obtained by interpolating the PIV data, representing the streamline direction at the intersection with the sampling circle. The dot product of the velocity vector $\textbf{u}(t)$ and the tangential vector $\textbf{r}'(\theta)$ is then calculated, and the optimal $\theta$ is identified by finding the minimum absolute value of this dot product.
\begin{equation}
    \alpha_{ins,circ}(t) = \arg \min_{\theta} \left(|\textbf{u}(t) \cdot \textbf{r}'(\theta)|\right)
\end{equation}

When the geometry is properly chosen, both instantaneous angle-of-attack sampling methods yield nearly identical results. In the discussions, we focus on the results obtained using the streamline interception method, selected for its simplicity and robustness.

\subsubsection*{The effect of instantaneous angle of attack sampling geometry}
\begin{figure}
\subfigure[$C_{L,w}$ deviates with $d$]{
    \includegraphics[width= 0.5\textwidth]{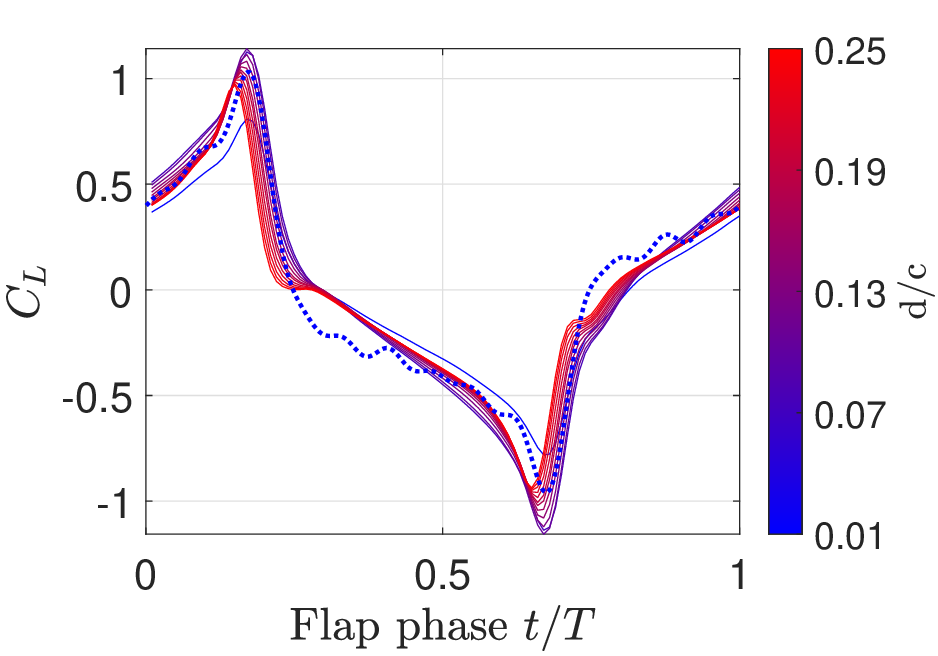}
    }
\subfigure[$C_{L,w}$ deviates with $h$]{
    \includegraphics[width= 0.5\textwidth]{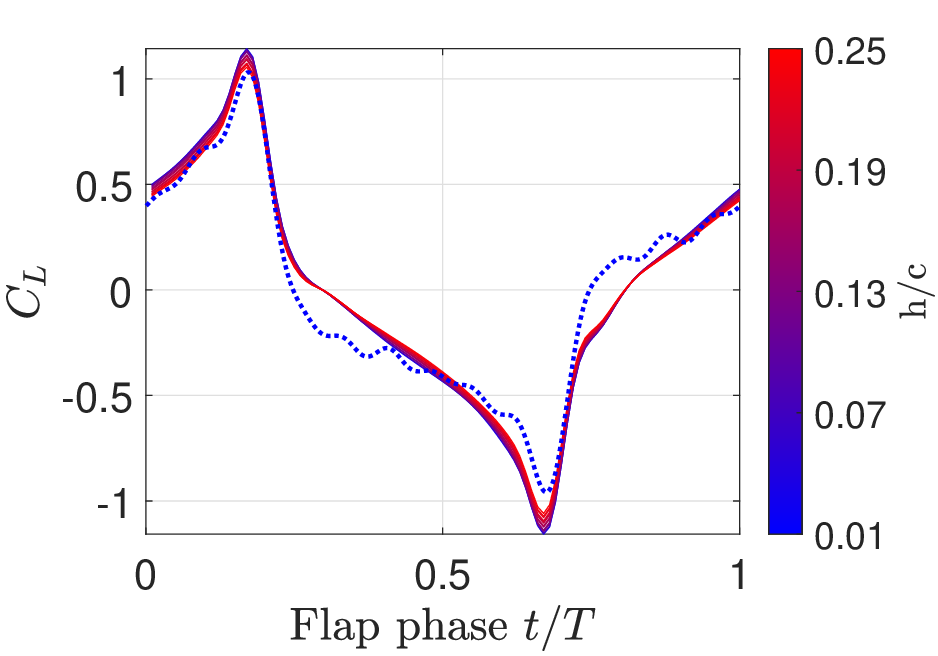}
    }
\subfigure[$C_{L,w}$ deviates with $w$]{
    \includegraphics[width= 0.5\textwidth]{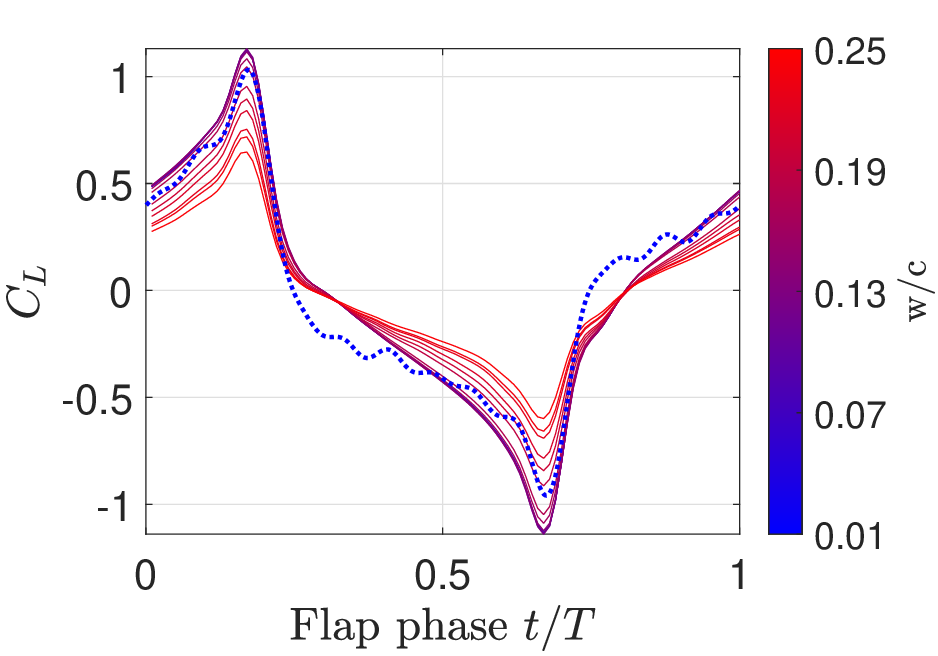}
    }

\caption{Deviation of $C_{L,w}$ with different sampling geometries and comparison with the directly measured lift coefficient $C_L$ (shown as a blue dashed line) in the case of St = 0.05, A/c = 0.5 with varying sampling geometries using the area averaging method, around $d = 0.05c$, $h = 0.12c$, and $w = 0.05c$.
}
\label{fig:geo_dependency1}
\end{figure}
The effect of each sampling geometry component on the prediction results is shown in figure~\ref{fig:geo_dependency1}. The deviation of $C_{L,w}$ is illustrated with varying sampling geometries around $d = 0.05c$, $h = 0.12c$, and $w = 0.05c$.

In figure~\ref{fig:geo_dependency1}a, when the sampling area is too close to the leading edge ($d < 0.03c$), the prediction loses accuracy due to reduced resolution of the PIV data near the wall, caused by a combination of laser reflection and near-wall stagnation. As $d$ increases, the prediction accuracy quickly improves; however, a lag emerges as the sampling area moves further from the leading edge.

As shown in figure~\ref{fig:geo_dependency1}b, the prediction is relatively insensitive to height variations of the rectangular sampling area, ranging from $0.01c$ to $0.25c$.

In figure~\ref{fig:geo_dependency1}c, the predicted lift amplitude decreases as the sampling width increases, due to the greater velocity field variance in the streamwise direction. Consequently, the instantaneous upwash and downwash effects are damped by wider width-direction averaging.

\begin{figure}
    \subfigure[$C_{L,w}$ deviates with $r$]{
    \includegraphics[width= 0.5\textwidth]{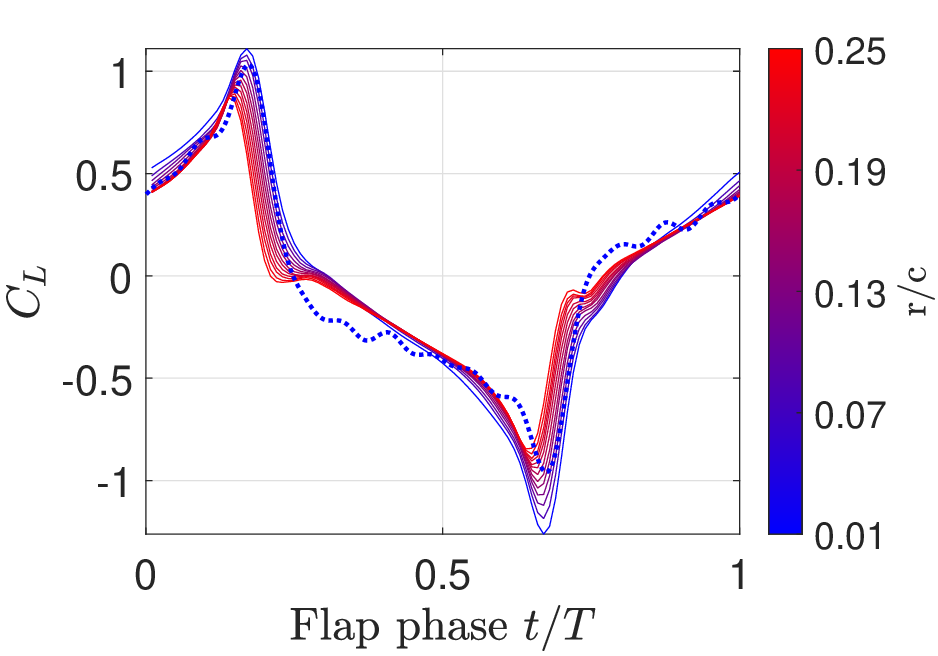}
    }
    \includegraphics[width= 0.5\textwidth]{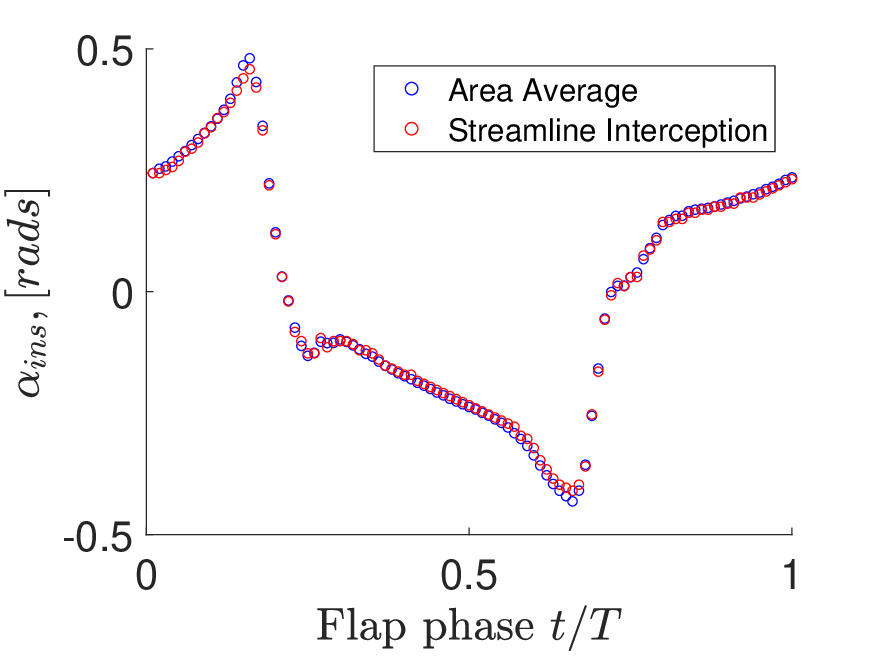}
\caption{(a) Deviation of the lift coefficient, $C_{L,w}$, calculated using the Wagner approach with varying radius values using the streamline interception method, from $0.01c$ to $0.25c$, compared to the directly measured lift coefficient, $C_L$ (blue dashed line), for $\mathrm{St} = 0.05$ and $A/c = 0.5$.   (b) Comparison of the result of area average and streamline interception method, with the base geometry parameters of  $d = 0.05c$, $h = 0.12c$, $w = 0.05c$ and $r=0.05c$.
}
\label{fig:geo_dependency2}
\end{figure}

As shown in figure~\ref{fig:geo_dependency2}a, the deviation of $C_{L,w}$ with different sampling radius values, using the streamline interception method, ranging from 0.01c to 0.25c, is compared with the directly measured lift coefficient $C_L$ (shown as a blue dashed line) for the case of St = 0.05 and A/c = 0.5. Both amplitude and phase shifts are observed with changes in the sampling geometry; however, the predicted lift coefficient remains relatively robust across a wide range of sampling circle radius $r$. The following results reported are all based on the streamline interception method with a sampling circle radius $r=0.05c$.

The instantaneous angle of attack acquired by the two sampling methods is compared in figure~\ref{fig:geo_dependency2}b using intermediate geometry parameters: $d = 0.05c$, $h = 0.12c$, $w = 0.05c$ for the area averaging approach, and $r = 0.05c$ for the streamline interception approach. The overall difference between the two results is subtle, with the streamline interception approach producing a slightly smoother profile at the turning points. Given the simplicity of the geometry, we find the streamline interception method to be both practical and robust, and have therefore adopted it as the basis for determining the instantaneous angle of attack in the discussion.
\subsection*{Duhamel integral}
\label{DuhamelApp}
The response to a step function $\mathbf{1}(t)$ is defined as the indicial admittance  \citep{fung_introduction_2008}: $A(t)$. For a step function applied at time $\tau$, the corresponding indicial admittance function is given by $A(t-\tau)$. In the case of a linear system, the total admittance can be superposed as the response to a combination of step functions, allowing for the construction of a time-varying response based on the system's linearity.
\begin{equation}
c_1\mathbf{1}(t)+c_2\mathbf{1}(t-\tau)\rightarrow c_1A(t)+c_2A(t-\tau)
\end{equation} 
For a derivable input function $f(t)$, it can always be deformed as 
\begin{equation}
f(t) = f(0)+\int_0^t \frac{df}{dt}(\tau) \mathbf{1}(t-\tau)d\tau
\end{equation}
Thus the corresponding admittance function is
\begin{equation}
x(t) = f(0)A(\tau)+\int_0^t \frac{df}{dt}(\tau) A(t-\tau)d\tau
\label{Duhamel}
\end{equation}
Eqn.~\ref{Duhamel} is known as the Duhamel integral of $f(t)$, sometimes also referred to as convolutional integration. In classic unsteady aerodynamics, where the system is considered linear according to thin airfoil theory, this integration is used to extend admittance functions, such as the Wagner or K\"ussner function, to time-varying flow conditions.

\subsection*{Phase and amplitude errors analysis of unsteady prediction approaches}

\begin{figure}
\subfigure[ ]{
    \includegraphics[width= 0.5\textwidth]{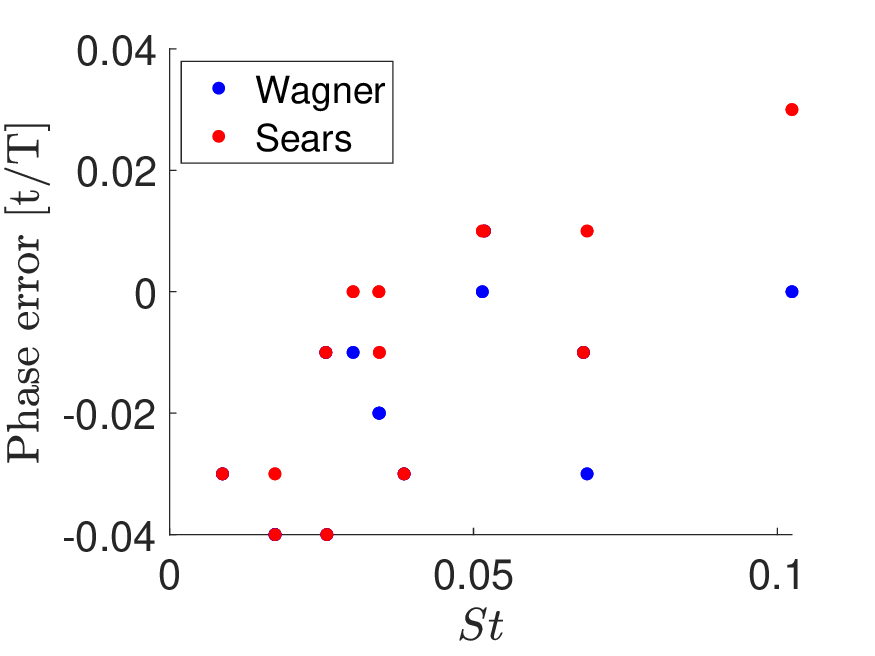}
    }
\subfigure[ ]{
    \includegraphics[width= 0.5\textwidth]{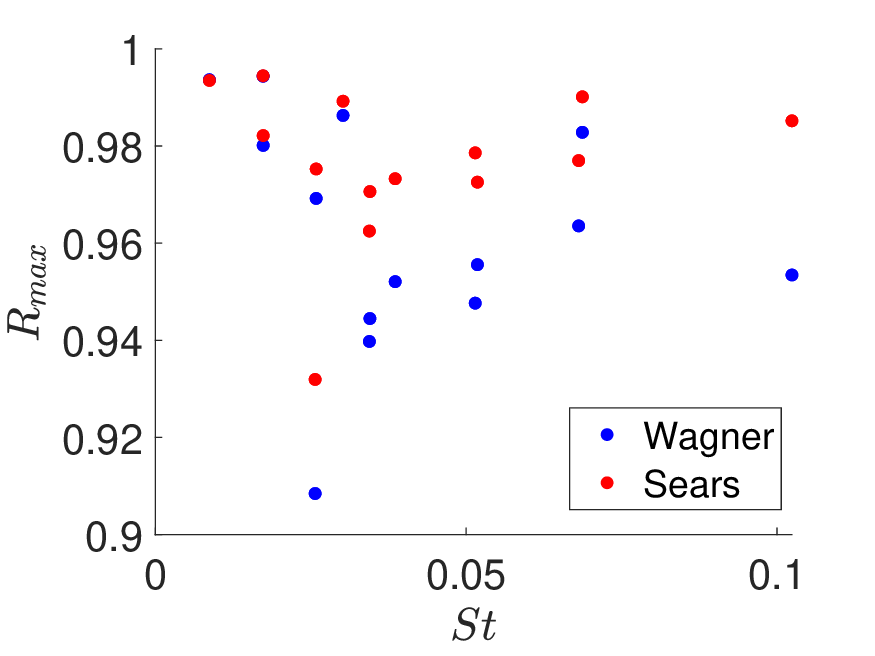}
    }

\caption{ 
Phase and Amplitude Errors Analysis
are presented: (a) show the phase lag  while (b) show $R_{max}$ for the Wagner (blue) and Sears (red) approach. 
}
\label{fig:error_corr}
\end{figure}

To further examine the phase and amplitude accuracy of the time and frequency approaches, the predicted lift profiles are compared with direct measurements using cross-correlation across different flow conditions, distinct by $St$, case by case. A cross-correlation is calculated between the predicted and measured lift coefficients:
\begin{equation}
    R_{C_L}(t/T) = C_{L,calc}(t/T) \star C_{L,meas}(t/T).
\end{equation}
The normalized correlation value is represented as a double-sided function of phase steps (based on phase-averaged flow conditions).  The phase error is identified by the position of the maximum correlation value,
\begin{equation}
    \mbox{phase error} = arg \max_{t/T}\{ R_{C_L} \}, 
\end{equation}  
while the amplitude accuracy is indicated by the magnitude of the maximum correlation value:
\begin{equation}
    R_{max} =\max \{ R_{C_L} \} .
\end{equation} 
The phase error and amplitude accuracy are determined through cross-correlation between the calculated and measured lift coefficients for both Wagner's approach and Sear's approach, shown in figure~\ref{fig:error_corr} (a) for phase lag and (b) for amplitude error.
The phase error for both approaches increases with St, transitioning from slightly negative to slightly positive. In terms of amplitude, Wagner approach generally demonstrates higher accuracy.

\subsection*{The convergence of unsteady prediction approaches }
\begin{figure}
\subfigure[ ]{
    \includegraphics[width= 0.5\textwidth]{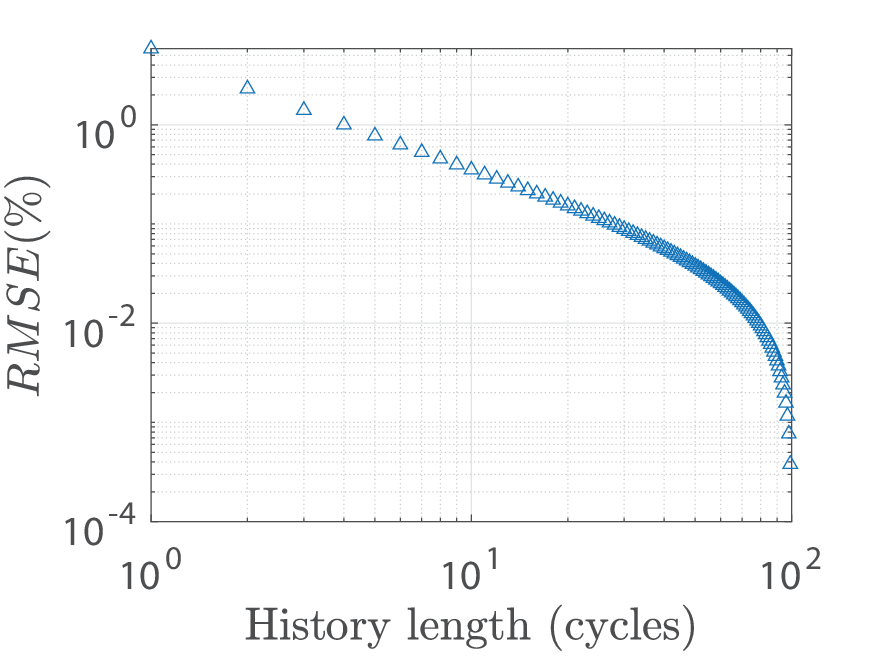}
    }
\subfigure[ ]{
    \includegraphics[width= 0.5\textwidth]{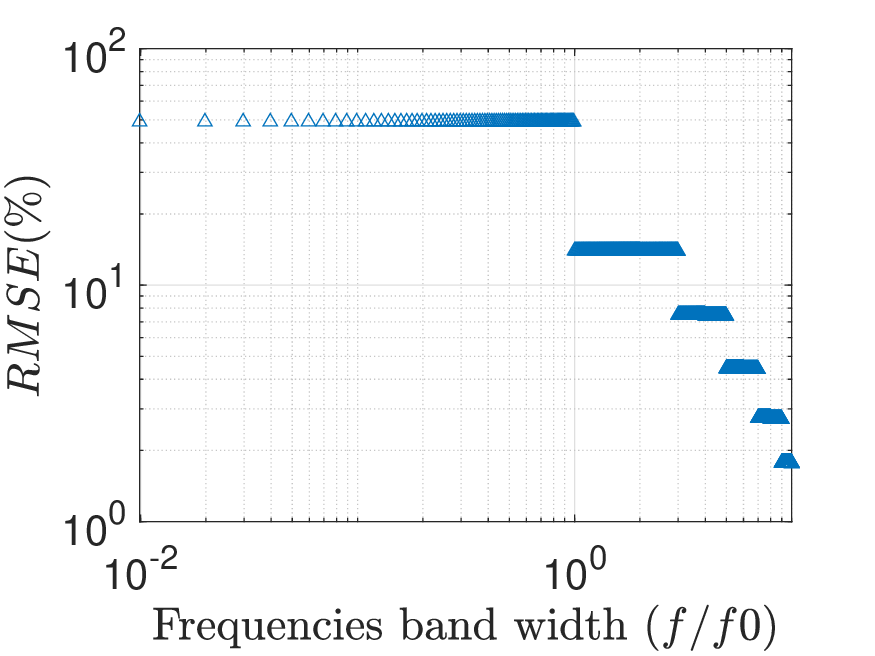}
    }
\caption{Convergence speed of the predictions (a) RMSE of the convolutional method (with Wagner function) as a function of time history length included in the calculation comparing with the final result with 100 cycles
(b) RMSE of the Fourier synthesis method (Sears's function) as a function of frequency band width included in the calculation comparing with the final result.
}
\label{fig:conv}
\end{figure}

Note that the result of the convolution method (the Duhamel integral \ref{Duhamel}) in the time domain has a dependency on the length of the time history included in the calculation.
The time-domain convergence of the convolution method (using the Wagner function), illustrated in figure~\ref{fig:conv}a, quantifies the impact of the time-history length on the results. These calculations are performed using the Duhamel integral while varying the number of cycles included in the history.
To measure convergence, an RMS error is defined relative to a reference solution obtained using 100 cycles of historical data—long enough to ensure convergence:
\begin{equation}
    RMSE_{cyci} = \frac{\sqrt{\sum_{j=1}^n (C_{L,w,cyci,j} - C_{L,w,cyc100,j})^2 / n}}{C_{L,max}},
\end{equation}
where \( j \) indexes the discrete phases within one cycle, and \( n \) is the total number of discrete phases. The RMS error, normalized by the maximum computed lift coefficient \( C_{L,max} \), provides a measure of the accuracy achieved with a given history length. By comparing these errors for different numbers of included cycles, we can evaluate how quickly the solution converges.

It is unsurprising that such a comparison converges, as the input from distant history becomes negligible eventually physically, also as a nature of the Duhamel integral mathematically. 
The key is that, however, the error level drops quickly within the first few cycles and falls below 0.1\% of $C_{L,max}$. This provides a convenient guideline for determining the appropriate length of time history to include in Duhamel integral.

Similarly, the results of the Fourier synthesis method using Sears' function in the frequency domain depend on the range of frequency components included in the calculation. As the included bandwidth (starting from zero expressed as \( f/f_0 \)) increases, the prediction converges toward the final result. This dependency can be quantified similarly to the time-domain approach, as shown in figure~\ref{fig:conv}b.
The results are computed using Fourier synthesis over a range of bandwidths, from 0 to \(10 f/f_0 \). An RMS error as a function of bandwidth is defined as:
\begin{equation}
    RMSE_{bandi} = \frac{\sqrt{\sum_{j=1}^n (C_{L,s,bandi,j} - C_{L,s,band100,j})^2 / n}}{C_{L,max}},
\end{equation}
where \( bandi \) represents the included bandwidth, \( j \) is the index of each discrete phase in one cycle, and \( n \) is the total number of discrete phases. The reference solution is obtained using a bandwidth of 100 times the driving frequency, which is sufficient to ensure convergence.

The error level starts quite high, as the information contained in the low-frequency range is limited. As the frequency band crosses the driving frequency ($f/f_0 = 1$), a sudden drop occurs since the energy peak is concentrated in the narrow band around $f/f_0 = 1$. A further reduction in error takes place as more features are recovered by increasing the bandwidth to cover the harmonic peaks. Notably, larger drops are observed at the odd-numbered harmonics (1, 3, 5, ...) compared to the even-numbered harmonics (2, 4, 6...). This feature consist with our observation in section \ref{cycle-lift} as well as  \citet{turhan_coherence_2022}, which shows that more of the fluctuation energy are located in the odd number harmonics. However, these observations are from the force responses. Since the force prediction is purely flow-based (from the instantaneous angle of attack), further confirms the feature is flow induced.

A natural thought would be to compare the different prediction results with the data measured directly in this convergence analysis. Given that there will be no exact match between the measured and predicted results, the comparison will converge to a certain value. However, the final converged value is not particularly meaningful, as the objective is not to quantify the accuracy but rather to assess the convergence speed and provide a guideline for the calculation.

\clearpage
\raggedright
\bibliographystyle{jfm}
\bibliography{referencesZ}

\end{document}